\documentclass[aps,prd,nofootinbib,showkeys,preprint,floatfix]{revtex4}
\pdfoutput=1

\usepackage{amssymb}
\usepackage{amsmath}
\usepackage{amsfonts}
\usepackage{graphicx}
\usepackage{color}
\usepackage{xspace}
\usepackage{ulem}
\usepackage{lscape}
\usepackage{wasysym}
\usepackage{mathrsfs} 

\usepackage{multirow,rotating}

\def\gsim{\raise0.3ex\hbox{$\;>$\kern-0.75em\raise-1.1ex\hbox{$\sim\;$}}}
\def\lsim{\raise0.3ex\hbox{$\;<$\kern-0.75em\raise-1.1ex\hbox{$\sim\;$}}}

\def\znbb{0\nu\beta\beta}

\newcommand{\ba}[1]{\begin{eqnarray} \label{(#1)}}
\newcommand{\ea}{\end{eqnarray}}
\newcommand{\rf}[1]{(\ref{(#1)})}

\newcommand{\AddrAHEP}{
  {\it AHEP Group, Instituto de F\'{\i}sica Corpuscular --
    C.S.I.C./Universitat de Val{\`e}ncia \\
    Edificio de Institutos de Paterna, Apartado 22085,
  E--46071 Val{\`e}ncia, Spain}}

  \newcommand{\AddrUFSM}{
Departamento de F\' isica, Facultad de Ciencias, Universidad de La Serena, \\
Avenida Cisternas 1200, La Serena, Chile.  \\
Centro-Cient\'\i fico-Tecnol\'{o}gico de Valpara\'\i so, \\ 
Casilla 110-V, Valpara\'\i so,  Chile.
}

\def\gsim{\raise0.3ex\hbox{$\;>$\kern-0.75em\raise-1.1ex\hbox{$\sim\;$}}}
\def\lsim{\raise0.3ex\hbox{$\;<$\kern-0.75em\raise-1.1ex\hbox{$\sim\;$}}}

\newcommand{\nn}{\nonumber}

\begin{document}

\preprint{IFIC/17-36}

\title{Lepton number violating phenomenology of $d=7$ neutrino mass models}

\author{R. Cepedello}\email{ricepe@ific.uv.es}\affiliation{\AddrAHEP}
\author{J.C. Helo} \email{jchelo@userena.cl}\affiliation{\AddrUFSM}
\author{M. Hirsch} \email{mahirsch@ific.uv.es}\affiliation{\AddrAHEP}



\begin{abstract}

  We study the phenomenology of $d=7$ 1-loop neutrino mass models. All
  models in this particular class require the existence of several new
  $SU(2)_L$ multiplets, both scalar and fermionic, and thus predict a
  rich phenomenology at the LHC. The observed neutrino masses and
  mixings can easily be fitted in these models. Interestingly, despite
  the smallness of the observed neutrino masses, some particular
  lepton number violating (LNV) final states can arise with observable
  branching ratios. These LNV final states consists of leptons and
  gauge bosons with high multiplicities, such as $4l+4W$, $6l+2W$
  etc. We study current constraints on these models from upper bounds
  on charged lepton flavour violating decays, existing lepton number
  conserving searches at the LHC and discuss possible future LNV
  searches.

\end{abstract}

\maketitle

\tableofcontents

\section{Introduction\label{Sect:Intro}}

A Majorana mass term for neutrinos always implies also the existence
of lepton number violating (LNV) processes. The best-known example is
neutrinoless double beta decay ($\znbb$), for reviews see
\cite{Deppisch:2012nb,Avignone:2007fu}.  A high-scale mechanism, such
as the classical seesaw type-I
\cite{Minkowski:1977sc,Yanagida:1979as,Mohapatra:1979ia}, however,
will leave no other LNV signal than $\znbb$ decay. From this point of
view, models in which the scale of LNV is around the electro-weak
scale are phenomenologically much more interesting.

Low-scale Majorana neutrino mass models need some suppression
mechanism to explain the observed smallness of neutrino masses.  (For
a recent review on theoretical aspects of neutrino masses see
\cite{Cai:2017jrq}.) This suppression could be due to loop factors
\cite{Bonnet:2012kz,Sierra:2014rxa}, or neutrino masses could be
generated by higher order operators \cite{Bonnet:2009ej,Babu:2009aq},
or both. In this paper, we will study the phenomenology of a
particular class of models, namely $d=7$ 1-loop models
\cite{Cepedello:2017eqf}. Our main motivation is that $d=7$ 1-loop
contributions to neutrino masses can be dominant only, if new
particles below approximately 2 TeV exist. This mass range can be
covered by the LHC experiments in the near future, if some dedicated
search for the LNV signals we discuss in this paper is carried out.

Lepton number violation has been searched for at the LHC so far using
the final state of same-sign dileptons plus jets, $l^{\pm}l^{\pm}jj$.
Many different LNV extensions of the standard model (SM) can lead to
this signal \cite{Helo:2013dla,Helo:2013ika}. However, ATLAS and CMS
searches usually concentrate on only two theoretical scenarios,
left-right symmetry \cite{Keung:1983uu} and the standard model
extended with ``sterile neutrinos''. Note that these two models lead
to the same final state signal, but rather different kinematical
regions are explored in the corresponding experimental searches. CMS
has published first results from searches at run-II \cite{CMS:2017uoz}
and run-I \cite{Khachatryan:2014dka}, both for $eejj$ and $\mu\mu jj$
final states, concentrating on the left-right symmetric model.
\footnote{CMS has searched also for $\tau\tau jj$
  \cite{Sirunyan:2017yrk}.  However, that search is not a test for
  LNV, since one $\tau$ is assumed to decay hadronically.}  There is
also a CMS search for sterile Majorana neutrinos, based on ${\cal L}=
19.7/fb$ at $\sqrt{s}=8$ TeV \cite{Khachatryan:2016olu}.  ATLAS
published a search for $lljj$ based on 8 TeV data, for both SM with
steriles and for the LR model \cite{Aad:2015xaa}.  However, only
like-sign lepton data was analyzed in \cite{Aad:2015xaa} and no update
for $\sqrt{s}=13$ TeV has been published so far from ATLAS. No signal
has been seen in any of these searches so far and thus lower (upper)
limits on masses (mixing angles) have been derived.

Other final states that can test LNV have been discussed in the
literature. For example, in the seesaw type-II \cite{Schechter:1980gr}
the doubly charged component of the scalar triplet $\Delta$ can decay
to either $\Delta^{++}\to l^+l^+$ or $\Delta^{++}\to W^+W^+$ final
states.  If the branching ratios to both of these final states are of
similar order, LNV can be established experimentally
\cite{Azuelos:2004mwa,Perez:2008ha,Melfo:2011nx}. No such search has
been carried out by the  LHC experiments so far.  Instead, ATLAS
\cite{ATLAS:2014kca,ATLAS:2016pbt,ATLAS:2017iqw} and
CMS \cite{CMS:2016cpz} have searched for invariant mass peaks in the
same-sign dilepton distributions. Assuming that the branching ratios
for $ee$ and/or $\mu\mu$ are large, i.e. ${\cal O}(1)$, lower limits
on the mass of the $\Delta^{\pm\pm}$ up to 850 GeV
\cite{ATLAS:2017iqw}, depending on the flavour, have been derived.
Note that, if only one of the two channels are observed, LNV can not
be established at the LHC but the type of scalar multiplet could be
still determined \cite{delAguila:2013yaa}.

Dimension-7 ($d=7$) neutrino mass models can lead to new LNV final
states at the LHC. The proto-type tree-level model of this kind has
been discussed first in \cite{Babu:2009aq}, in the following called
the BNT model. As pointed out in \cite{Babu:2009aq} the model predicts
the final state $W^{\pm}W^{\pm}W^{\pm}+W^{\mp}l^{\mp}l^{\mp}$. The LHC
phenomenology of the BNT model has been studied recently in detail in
\cite{Ghosh:2017jbw}. Again, as in the case of
$W^{\pm}W^{\pm}+l^{\mp}l^{\mp}$ predicted by the seesaw type-II, no
experimental search for this particular LNV final state has been
published so far.

At tree-level the BNT model is unique in the sense that it is the only
$d=7$ model that avoids the lowest order $d=5$ contribution to the
neutrino mass, without relying on additional (discrete) symmetries
\cite{Bonnet:2009ej,Cepedello:2017eqf}.  Recently, we have studied
systematically $d=7$ 1-loop neutrino mass models
\cite{Cepedello:2017eqf}. These models, while necessarily more rich in
their particle content than simple $d=5$ (or $d=7$) tree-level
neutrino mass models, offer a variety of interesting LNV signals at
the LHC, so far not discussed in the literature. As we show below,
depending on the unknown mass spectrum, several different multi-lepton
final states with gauge bosons up to
$W^{\pm}W^{\pm}l^{\mp}l^{\mp}+l^{\pm}l^{\pm}l^{\mp}l^{\mp}$ can occur.
Note that for such high multiplicity final states one can expect very
low SM backgrounds.

Apart from LNV signals, the parameter space of $d=7$ neutrino mass
models can be constrained by a variety of searches. First, neutrino
masses and angles should be correctly fitted. Since we now know that
all three active neutrino mixing angles are non-zero, this fit leads
to certain predictions for lepton flavour violating decays.  We
therefore discuss also current and future constraints coming from
$\mu\to e \gamma$, $\mu \to 3 e$ and $\mu\to e$-conversion in nuclei.

Constraints on our models come also from lepton number conserving LHC
searches.  The same-sign dilepton searches
\cite{ATLAS:2014kca,ATLAS:2016pbt,ATLAS:2017iqw,CMS:2016cpz},
discussed above, can be recasted into lower mass limits valid for our
models. In addtion, also multi-lepton searches \cite{Sirunyan:2017qkz},
motivated by the seesaw type-III, can be used to obtain interesting
limits. We note in passing that we have also checked that the LNV
searches for $lljj$ \cite{Aad:2015xaa,CMS:2017uoz} are currently not
competitive for the models we consider in this paper.

The rest of this paper is therefore organized as follows. In the next
section, we discuss the basic setup of $d=7$ models and then present
the Lagrangians of our two example models. Section
\ref{Sect:Constraints} then calculates neutrino masses and constraints
from low energy probes.  Section \ref{Setc:LNV} discusses LHC
phenomenology. We first derive constraints from existing searches,
before discussing possible searches for LNV final state. We then close
with a short summary and discussion.

\section{Theoretical setup: $d=7$ models\label{Sect:Dim7}}

\subsection{$d=7$ neutrino mass models}

Before we discuss our example models, it may be useful to recapitulate
some basics about Majorana neutrino masses in general and $d=7$
models in particular. Majorana neutrino masses can be generated 
from $d=5+2n$ operators:
\begin{equation}\label{eq:d2n}
  {\cal O}^{d=5+2n}= LLHH \times ({HH^{\dagger}})^n
\end{equation}
The lowest order, $d=5$, is the well-known Weinberg operator
\cite{Weinberg:1979sa}. At tree-level, the Weinberg operator
has three types of ultra-violet completions \cite{Ma:1998dn},
known in the literature as seesaw type-I, type-II and type-III.
These (simplest) neutrino mass models make use of either
a right-handed neutrino (type-I), a scalar triplet (type-II)
or a fermionic triplet with zero hypercharge (type-III).

Higher order contributions to neutrino masses are expected to be
subdominant, unless the underlying model does not generate ${\cal
  O}^W$.\footnote{${\cal O}^W$ and higher order operators could give
  similar contibutions to neutrino masses, if the coefficient of
  ${\cal O}^W$ is small. We are not interested in this case.}  This
can be achieved essentially in two ways: Either via introducing a
discrete symmetry \cite{Bonnet:2009ej} or simply because the particle
content of the model does not allow to complete the lowest order
operator \cite{Babu:2009aq,Cepedello:2017eqf}. We will not be
interested in models with additional discrete symmetries here, since
such models, although interesting theoretically, usually are based on
additional SM singlet states, which leave very little LHC
phenomenology to explore.\footnote{``Sterile'' neutrino searches at
  the LHC, see introduction, provide of course constraints on these
  models.} Consider, instead, the BNT model \cite{Babu:2009aq}. This
$d=7$ tree-level model introduces a vector-like fermion pair, $\Psi$
and ${\bar\Psi}$ with quantum numbers ${\bf 3}^F_1$ and a scalar
quadruplet $S \equiv {\bf 4}^S_{3/2}$. (Here and elsewhere we will 
use a notation which gives the $SU(2)_L$ representation
and hypercharge in the form ${\bf R}_Y$ with a superscript $S$ or $F$,
where necessary.) By construction, at tree-level
the lowest order contribution to the neutrino masses is $d=7$, see
fig. (\ref{fig:treed7}). Being higher order, already at tree-level,
two new particles are needed in order to generate a neutrino mass.
This model has a rich LHC phenomenology
\cite{Babu:2009aq,Ghosh:2017jbw} and, in particular, generates the LNV
final state $W^{\pm}W^{\pm}W^{\pm}+W^{\mp}l^{\mp}l^{\mp}$.

\begin{center}
\begin{figure}[tbph]
\begin{centering}
\includegraphics[scale=1.0]{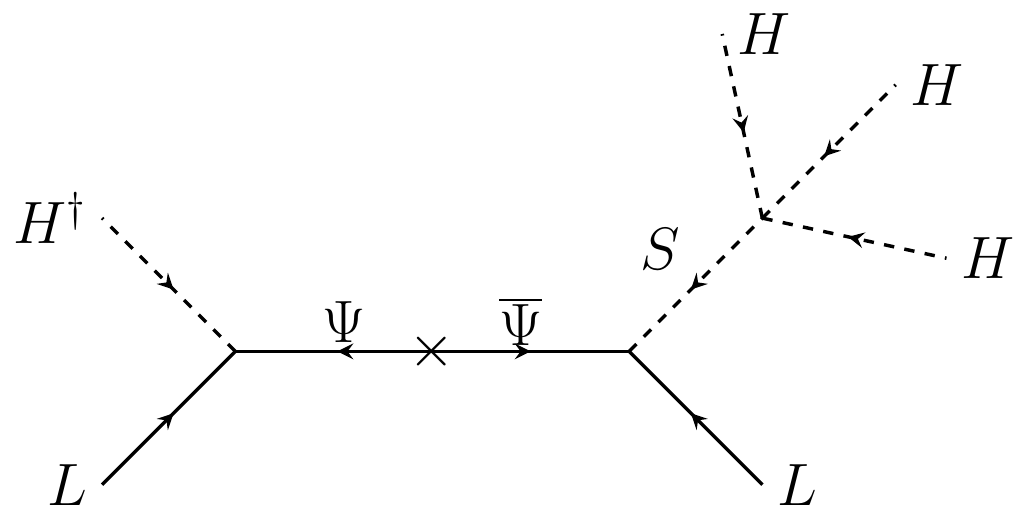}
\end{centering}
\protect\caption{\label{fig:treed7} $d=7$ neutrino 
mass diagram, for the BNT model  \cite{Babu:2009aq}. }
\end{figure}
\par
\end{center}
As mentioned in the introduction, the BNT model is unique at
tree-level in the sense that no additional symmetries are required to
make it the leading contribution to neutrino masses (we call such
models ``genuine''). In a recent paper \cite{Cepedello:2017eqf}, we
have analyzed systematically $d=7$ 1-loop models. While there exists a
large number of topologies, only a few of them can lead to genuine
$d=7$ models. These topologies can still generate 23 different
diagrams, but all models underlying these diagrams share the following
common features: (i) five new multiplets must be added to the SM
particle content; and (ii) all models contain highly charged
particles. In all cases there is at least one triply charged
state. Thus, see also the discussion, one expects that all $d=7$
1-loop models have rather similar accelerator phenomenology. For this
reason, in this paper we concentrate on only two of the simplest
example models.\footnote{Strictly speaking this is true only for
  variants of the $d=7$ 1-loop models for which the particles
  appearing in the loop are colour singlets. For a brief discussion
  for the case of coloured particles see section \ref{Sect:Con}.}

According to \cite{Cepedello:2017eqf} one can classify the $d=7$
models w.r.t. increasing size of the largest $SU(2)_L$ multiplet. 
There is one model, in which no representation larger than triplets
is needed. All other models require at least one quadruplet. Our
two example models, introduced below, are therefore just the
simplest realizations of ${\cal O}^{d=7}$ at 1-loop, but are expected
to cover most of the interesting phenomenology.

Finally, let us mention that the $d=7$ operator, see
eq. (\ref{eq:d2n}), generates automatically also a 1-loop $d=5$
neutrino mass:
\begin{equation}\label{eq:nlp1}
\frac{1}{\Lambda^3}LLHHHH^{\dagger} \rightarrow \frac{1}{16 \pi^2} 
\frac{1}{\Lambda}LLHH
\end{equation}
It is easy to estimate that this loop contribution will become more
important than the tree-level if $(\Lambda/v) \gsim 2$ TeV. Our main
motivation for the present study is that the LHC can explore large
parts of this parameter space.

\subsection{Triplet model}

Our first example model is the ``minimal'' 1-loop $d=7$ model. This
model is minimal in the sense that it uses no multiplet larger than
triplets. The model adds two new (vector-like) fermions and three
scalars to the standard model particle content:
\begin{center}
\begin{tabular}{ c c c }

$\Psi=\left(
\begin{matrix}
\Psi^{++} \\ 
\Psi^{+}  \\
\Psi^{0}
\end{matrix} 
\right) \sim \textbf{3}_1^F$

&\qquad

$\eta_1=\left(
\begin{matrix}
\eta_1^{++} \\ 
\eta_1^{+}
\end{matrix} 
\right) \sim \textbf{2}_{3/2}^S$

&\qquad

$\eta_2=\left(
\begin{matrix}
\eta_2^{+++} \\ 
\eta_2^{++}
\end{matrix} 
\right) \sim \textbf{2}_{5/2}^S$

\end{tabular}

\vspace*{0.5cm}

\begin{tabular}{ c c }

$\eta_3=\left(
\begin{matrix}
\eta_3^{++++} \\ 
\eta_3^{+++}  \\
\eta_3^{++}
\end{matrix} 
\right) \sim \textbf{3}_3^S$

&\qquad

$\psi_1=\left(
\begin{matrix}
\psi_1^{+++} \\ 
\psi_1^{++}
\end{matrix} 
\right) \sim \textbf{2}_{5/2}^F$.

\end{tabular}
\end{center}
Note that both, $\Psi$ and ${\bar\Psi}$ are needed. The Lagrangian of
the model contains the following terms:
\ba{lagr_3plet}
\mathscr{L}
&=& \left[ Y_1 H^\dagger \Psi P_L L + Y_2 \overline{\psi}_1 P_L L \eta_3
  + Y_3 \eta_1^\dagger \overline{\Psi} \psi_1 + Y_4 \eta_1 \overline{\Psi} P_L L
  + Y_5\, e_R \eta_1^\dagger \chi_1 + \text{H.c.}
  \right] \\ \nn &-& M_\Psi \overline{\Psi}\Psi - M_{\psi_1}
\overline{\psi}_1 \psi_1 - V_{scalar} ,
\ea
with the scalar part given by:
\ba{pot_3plet}
V_{scalar} &=& m_H^2 H^\dagger H + m_{\eta_1}^2
\eta_1^\dagger \eta_1 + m_{\eta_2}^2 \eta_2^\dagger \eta_2 +
m_{\eta_3}^2 \eta_3^{\dagger} \eta_3 \\ \nn &+& \left[ \mu_1 \,
  H \eta_2 \eta_3^{\dagger} + \mu_2
  \eta_1 \eta_1 \eta_3^{\dagger} +
  \lambda_2 \eta_2^\dagger H \eta_1 H + \lambda_3 \, \eta_1^\dagger \eta_2 \eta_1^\dagger H +
  \text{H.c.} \right]  \\ \nn
  &+& \frac 12 \lambda_1 (H^\dagger H)^2 + \frac 12
\lambda_4 (\eta_1^\dagger \eta_1)^2 + \frac 12 \lambda_5
(\eta_2^\dagger \eta_2)^2 + \frac 12 \lambda_6 (\eta_3^{\dagger}
\eta_3)^2 \\ \nn &+& \lambda_7 (H^\dagger H)(\eta_1^\dagger \eta_1)
+ \lambda_8 (H^\dagger H)(\eta_2^\dagger \eta_2) + \lambda_9
(H^\dagger H)(\eta_3^{\dagger} \eta_3) + \lambda_{10}
(\eta_1^\dagger \eta_1)(\eta_2^\dagger \eta_2) \\ \nn &+& \lambda_{11}
(\eta_1^\dagger \eta_1)(\eta_3^{\dagger} \eta_3) + \lambda_{12}
(\eta_2^\dagger \eta_2)(\eta_3^{\dagger} \eta_3) + \lambda_{13}
(H^\dagger \eta_1)(\eta_1^\dagger H) + \lambda_{14} (H^\dagger
\eta_2)(\eta_2^\dagger H) \\ \nn &+& \lambda_{15} ( H^\dagger
\eta_3 )( \eta_3^{\dagger} H ) + \lambda_{16} (
\eta_1^{\dagger} \eta_2 )( \eta_2^\dagger \eta_1 ) + \lambda_{17} (
\eta_1^\dagger \eta_3 )( \eta_3^{\dagger} \eta_1
) + \lambda_{18} ( \eta_2^\dagger \eta_3 )(\eta_3^{\dagger}\eta_2 )
\ea
The model contains many charged scalars, but the only neutral scalar
is the standard model Higgs.

From the Yukawa couplings only $Y_1$, $Y_2$, $Y_3$ enter the neutrino
mass calculation directly, see next section.  Similarly, from the
scalar terms only the coupling $\lambda_2$ and mass term $\mu_1$ and
the mass matrix of the doubly charged scalars play an important role.
We therefore give here only the mass matrix for the $S_i^{++}$
states. In the basis ($\eta_1,\eta_2,\eta_3$) it is given as
\begin{equation}\label{eq:etpp}
{\cal M}_{\eta^{++}}^2 =
\begin{pmatrix}
m_{S_1}^2 & -\frac{\lambda_2 v^2}{2} & 0 \\
-\frac{\lambda_2 v^2}{2} & m_{S_2}^2 &  -\frac{\mu_1 v}{\sqrt{2}} \\
0 &   -\frac{\mu_1 v}{\sqrt{2}} &  m_{S_3}^2
\end{pmatrix} .
\end{equation}
Here, $v$ is the SM Higgs vacuum expectation value (vev) and: 
\begin{eqnarray}\label{eq:defmsq}
  m_{S_1}^2 = m_{\eta_1}^2 + \frac{\lambda_7}{2}v^2, \\ \nonumber
  m_{S_2}^2 = m_{\eta_2}^2 + \frac{\lambda_8+\lambda_{14}}{2}v^2, \\ \nonumber
  m_{S_3}^2 = m_{\eta_3}^2 + \frac{\lambda_9+\lambda_{15}}{2}v^2. \\ \nonumber
\end{eqnarray}
Eq.(\ref{eq:etpp}) can be diagonalized by
\begin{equation}\label{eq:diagetpp}
  {\hat{\cal M}}_{\eta^{++}}^2 = R_{\eta^{++}}^T {\cal M}_{\eta^{++}}^2 R_{\eta^{++}} .
\end{equation}
All other mass matrices of the model can be easily derived and we do not
give them here for brevity.

\subsection{Quadruplet model}

Our second example model makes use of the quadruplet  $S$.
The full new particle content of the model is:

\begin{center}
\begin{tabular}{ c c c }

$S=\left(
\begin{matrix}
S^{+++} \\
S^{++} \\ 
S^{+}  \\
S^{0}
\end{matrix} 
\right) \sim \textbf{4}_{3/2}^S$

\quad & \quad

$\chi_1=\left(
\begin{matrix}
\chi_1^{++} \\ 
\chi_1^{+}
\end{matrix} 
\right) \sim \textbf{2}_{3/2}^F$

\quad & \quad

$\chi_2=\left(
\begin{matrix}
\chi_2^{++++} \\ 
\chi_2^{+++} \\ 
\chi_2^{++}
\end{matrix} 
\right) \sim \textbf{3}_3^F$

\end{tabular}
\begin{tabular}{ c c }

$\phi_1=\phi_1^{++} \sim \textbf{1}_2^S$

\qquad & \qquad

$\phi_2=\left(
\begin{matrix}
\phi_2^{+++} \\ 
\phi_2^{++}
\end{matrix} 
\right) \sim \textbf{2}_{5/2}^S$.

\end{tabular}
\end{center}
Again, fermions need to be vector-like. The Lagrangian of the
model is given by:

\ba{lagr_4plet}
\mathscr{L} &=& \left[ Y_1 \overline{\chi}_1 P_L L \phi_1 + Y_2 \phi^\dagger_2 P_L L \chi_2 + Y_3 \chi_1 S\, \overline{\chi}_2 + Y_4 e_R \overline{\chi}_1 \phi_2 + Y_5 e_R H^\dagger \chi_1 \right.
	\\ \nn 
	&+& \left. Y_6 e_R e_R \phi_1 + H.c. \right] - M_{\chi_1} \overline{\chi}_1 \chi_1 - M_{\chi_2} \overline{\chi}_2 \chi_2
- V_{scalar}	,
\ea
with the scalar potential:
\ba{pot_4plet}
	V_{scalar} &=& m_H^2 H^\dagger H + m_S^2 S^\dagger S + m_{\phi_1}^2 \phi_1^\dagger \phi_1 + m_{\phi_2}^2 \phi_2^\dagger \phi_2
	\\ \nn 
	&+& \left[ \mu_1 \phi_1^\dagger H^\dagger \phi_2 + \lambda_2 S^\dagger H H H + \lambda_3 \phi_2^\dagger S H H + \lambda_4 \phi_2^\dagger S H^\dagger S +  \text{H.c.} \right] 
	\\ \nn 
	&+& \frac 1 2 \lambda_1 ( H^\dagger H )^2 + \frac 1 2 \lambda_5 ( \phi_1^\dagger \phi_1 )^2 + \frac 1 2 \lambda_6 ( \phi_2^{\dagger} \phi_2 )^2 + \frac 1 2 \lambda_7 ( S^\dagger S )^2 + \lambda_8 ( H^\dagger H )( \phi_1^\dagger \phi_1 ) 
	\\ \nn 
	&+& \lambda_9 ( H^\dagger H ) ( \phi_2^\dagger \phi_2 ) + \lambda_{10} ( H^\dagger H ) ( S^\dagger S ) + \lambda_{11} ( \phi_1^\dagger \phi_1 ) ( \phi_2^\dagger \phi_2 ) +  \lambda_{12} ( \phi_1^\dagger \phi_1 ) ( S^\dagger S )
	\\ \nn 
	&+& \lambda_{13} ( \phi_2^\dagger \phi_2 ) ( S^\dagger S ) + \lambda_{14} ( H^\dagger \phi_2 ) ( \phi_2^\dagger H ) + \lambda_{15} ( H^\dagger S )( S^\dagger H ) + \lambda_{16} ( S^\dagger \phi_2 )( \phi_2^\dagger S ).
\ea
Note that the term proportional to $\lambda_2$ will induce a non-zero
value for the vev of the neutral scalar $S$, even if $m_S^2$ is larger
than zero. One can thus take either $\lambda_2$ or $v_S$ as a free
parameter. In our numerical calculation we choose $v_S$, see below.

\section{Low energy constraints\label{Sect:Constraints}}

In this section we will discuss non-accelerator constraints on the
parameters of our two example models. We consider first neutrino
masses and angles and then turn to lepton flavour violating (LFV)
decays. The LHC phenomenology is discussed in section \ref{Setc:LNV}.

We have implemented both of our example models in SARAH
\cite{Staub:2012pb,Staub:2013tta}. Using Toolbox \cite{Staub:2011dp},
the implementation can be used to generate SPheno code
\cite{Porod:2003um,Porod:2011nf}, for the numerical evaluation of mass
spectra and observables, such as LFV decays ($\mu \to e \gamma$, $\mu
\to 3 e$ etc) calculated using Flavour Kit \cite{Porod:2014xia}. The
Toolbox subpackage SSP has then be used for our numerical scans.

\subsection{Neutrino masses}

\begin{center}
\begin{figure}[t]
\includegraphics[width=0.45\linewidth]{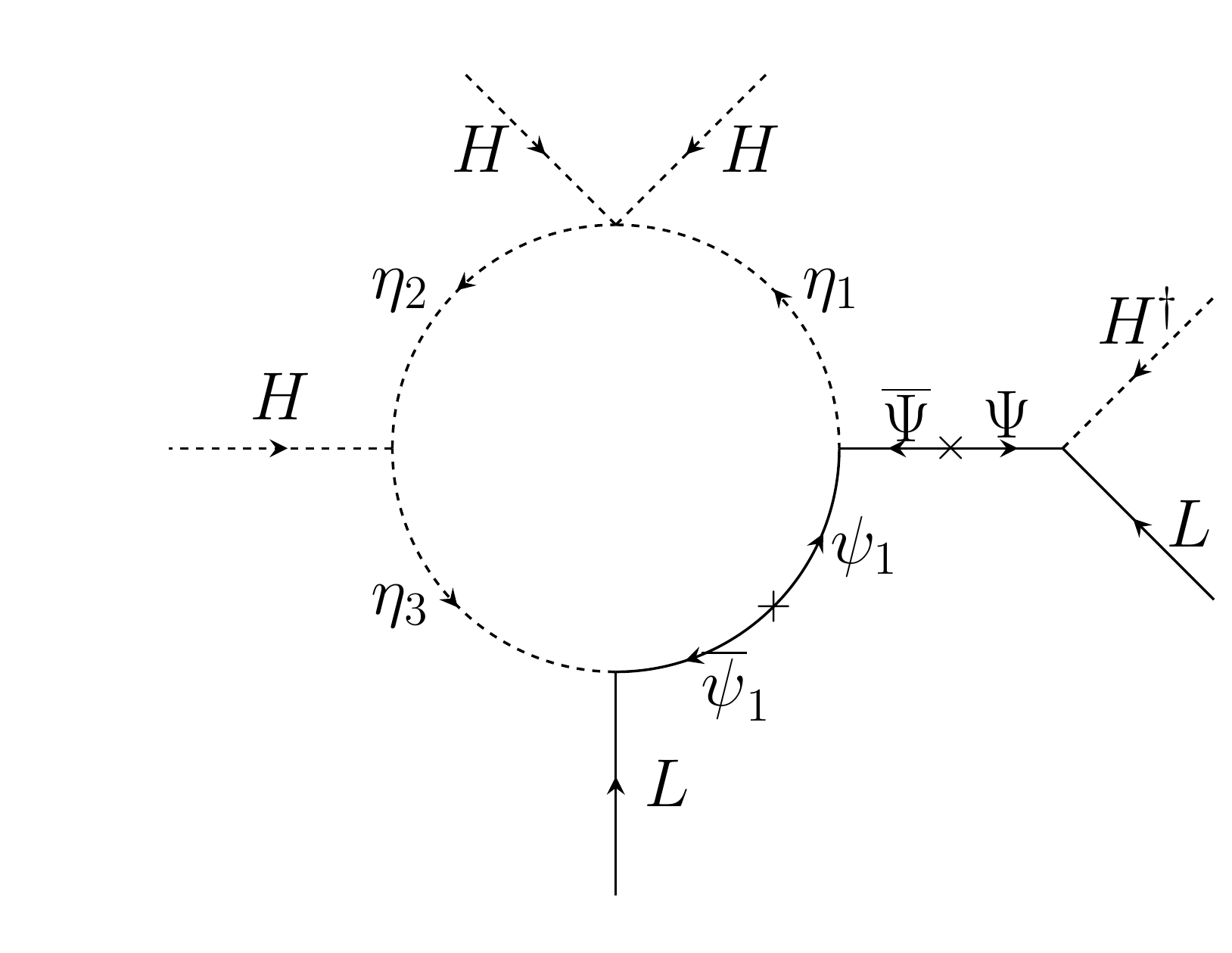}
\includegraphics[width=0.45\linewidth]{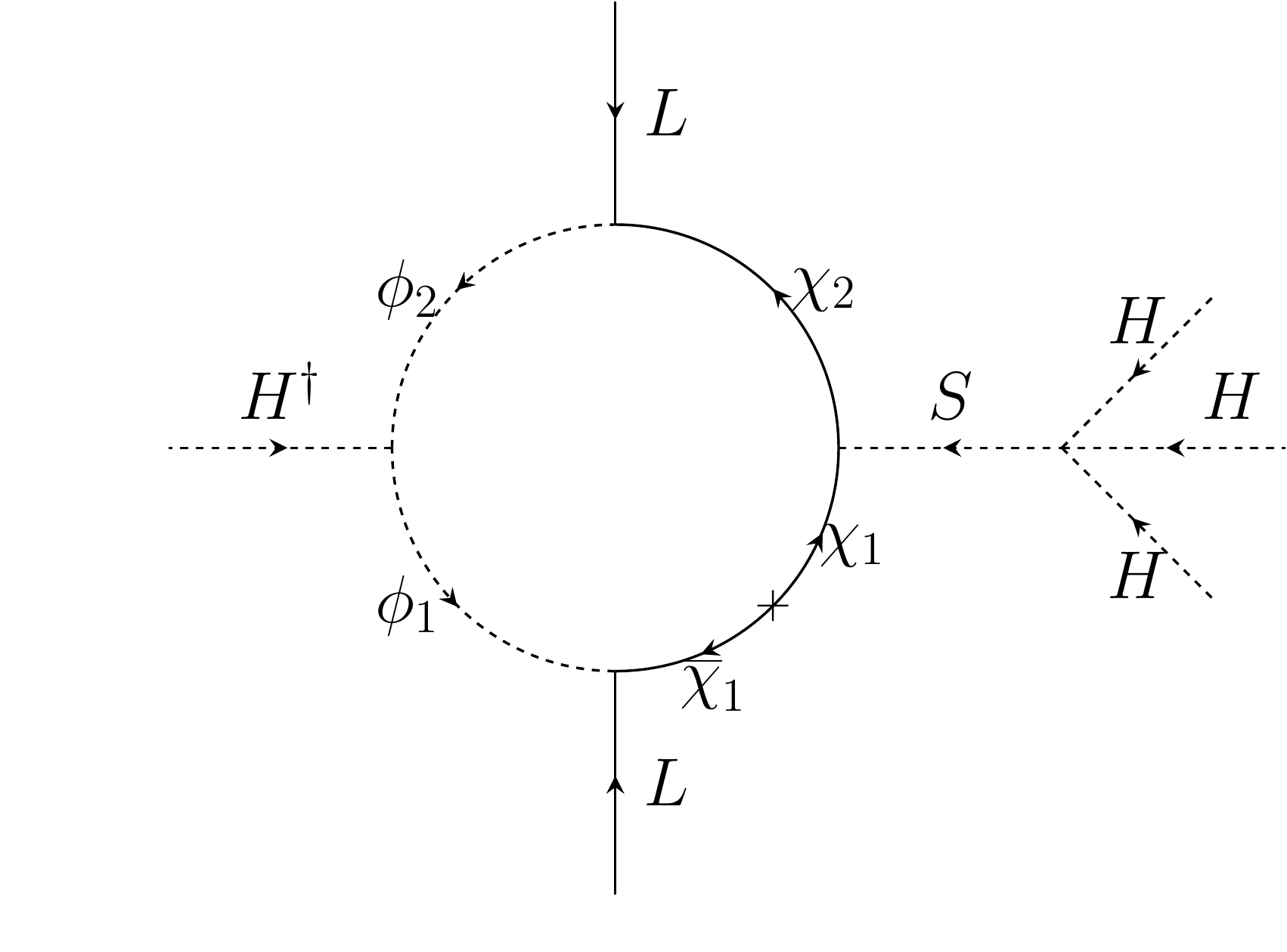}
\caption{1-loop neutrino mass diagrams for the triplet model (left)
  and for the quadruplet model (right). Diagrams are given in the
gauge basis. For a discussion see text.}
\label{fig:diag}
\end{figure}
\end{center}

Here we discuss the calculation of neutrino masses in our two example
models. We first consider the triplet model, then only briefly
summarize the calculation of the quadruplet model, since the
calculation is very similar in both cases. Note that SPheno allows to
calculate 1-loop corrected masses numerically. We have checked that
the description given below agrees very well with the numerical
results from SPheno.

The triplet model is described by the Lagrangian given in
eq. \rf{lagr_3plet} and generates d=7 1-loop neutrino masses via the
diagram shown in fig. (\ref{fig:diag}) to the left. Rotating the doubly
charged scalars to the mass eigenstate basis, the diagram in
fig. (\ref{fig:diag}) results in a neutrino mass matrix
given by:
\footnote{Eq. (\ref{eq:mnu1}) is already an approximation: $\Psi_0$
  mixes with the light active neutrinos. So, the total neutral fermion
  mass matrix is (4,4). However, this mixing should not be too large
  and is estimated here simply by the factor $\frac{Y_3 v}{m_\Psi}$.}
\begin{eqnarray}
  (m_{\nu})_{\alpha\beta}&=& \frac{1}{16 \pi^2} \frac{Y_3 v}{m_\Psi}m_{\psi_1}
  \sum_i (R_{\eta^{++}})_{1i} (R_{\eta^{++}})_{3i}  B_0(0, m^2_{\psi_1}, m^2_{S_i}) 
\left[
(Y_1)_{\alpha}
(Y_2)_{\beta}
+
(Y_1)_{\beta}
(Y_2)_{\alpha}
\right].
\label{eq:mnu1}
\end{eqnarray}
Here $(R_{\eta^{++}})$ is the rotation matrix defined in
eq. (\ref{eq:diagetpp}) and $m_{S_i}$ are the eigenvalues of
eq. (\ref{eq:etpp}).  $B_0(0, m^2_{\psi_1}, m^2_{S_i})$ is a
Passarino-Veltman function. In the numerical calculation we have used
eq. (\ref{eq:mnu1}) to fit the neutrino masses of the model to
neutrino oscillation data. However, in order to have a better
understanding of the dependence of eq. (\ref{eq:mnu1}) on the
different parameters of the Lagrangian, eq. \rf{lagr_3plet}, we also
give the expression of the neutrino mass matrix in the so-called mass
insertion approximation.  This approximation consists in replacing the
full diagonalization matrices and eigenvalues of the doubly charged
scalar mass matrix by their leading order ones. The resulting equation
can be written simply as:
\begin{eqnarray}
(m_{\nu})_{\alpha\beta}&=& {\cal F}
\times  \left[
(Y_1)_{\alpha}
(Y_2)_{\beta}
+
(Y_1)_{\beta}
(Y_2)_{\alpha}
\right],
\label{eq:mnu2}
\end{eqnarray}
where 
\begin{equation}
{\cal F}=  \frac{1}{16 \pi^2}
           \frac{Y_3 v}{m_\Psi}
\frac{v^2 \lambda_2}{m^2_{S_2}-m^2_{S_1}}
\frac{v \mu_1}{m^2_{S_3}-m^2_{S_2}} m_{\psi_1}
\left[\frac{m^2_{S_1}}{m^2_{\psi_1}-m^2_{S_1}}
  ln\left(\frac{m^2_{S_1}}{m^2_{\psi_1}}\right)
  - \frac{m^2_{S_2}}{m^2_{\psi_1}-m^2_{S_2}}
  ln\left(\frac{m^2_{S_2}}{m^2_{\psi_1}}\right)\right] 
\label{eq:prefac}
\end{equation}
Eq.(\ref{eq:mnu2}) shows that neutrino angles predicted by the model
depend on ratios of Yukawa couplings, while the overall mass scale is
determined by the prefactor ${\cal F}$. The model has the interesting
feature that $\det(m_{\nu})=0$.  Therefore it can fit only
hierarchical neutrino mass spectra (normal or inverse), but not a
degenerate spectrum \footnote{ In order to fit also a quasi-degenerate
  spectrum we would need to include more than one copy of $\Psi$ or/and $\psi_1$.}.
The eigenvalues of Eq.~\eqref{eq:mnu2} are:
\begin{equation}
m_{\nu_{1(3)}} = 0,
\quad 
m_{\nu_{2,3(1,2)}} = 
\left[
\sum_{\alpha} (Y_1)_{\alpha} (Y_2)_{\alpha} 
\mp 
\sqrt{
\sum_{\alpha}\left|(Y_1)_{\alpha}\right|^{2} 
\sum_{\alpha}\left|(Y_2)_{\alpha} \right|^{2}
}
\right]
{\cal F}
\label{eq:mnu-eigen}
\end{equation}
for normal (inverted) hierarchy.  From Eq.~\eqref{eq:mnu-eigen}, one
can estimate the constraints from neutrino masses on the size of the
Yukawa couplings.  In order to reproduce the neutrino mass suggested
by atmospheric neutrino oscillations ($m_{\nu_{3}} \sim 0.05$ eV),
keeping the mass scale of the new particles $M \sim$ 1 TeV, the scalar
coupling $\lambda_2 \sim 1$ and mass term $\mu \sim $ 1 TeV, the
Yukawa couplings $Y_{1}$, $Y_{2}$, $Y_{3}$ must be set typically to
$\mathcal{O}(10^{-2})$. Note, however, that this is only a rough estimate
and in our numerical calculations we scan over the free parameters
of the model. As discussed in the next subsection, LFV
produces upper limits on these Yukawa couplings very roughly of
this order.

In our numerical fits to neutrino data, we do not only fit to
solar and atmospheric neutrino mass differences, but also to
the observed neutrino angles \cite{Forero:2014bxa}. This is
done in the following way. First, we choose all free parameters
appearing in the prefactor ${\cal F}$. These leaves us with the
six free parameters in the two vectors $Y_1$ and $Y_2$. Two neutrino
masses and three neutrino angles give us five constraints. We
arbitrarily choose $(Y_1)_e$ as a free parameter, the remaining
five entries are then fixed. Since $\det(m_{\nu})=0$, finding the
solutions for those five parameters implies solving coupled quadratic
equations, which can be done numerically.

For the quadruplet model we show the neutrino mass diagram in
fig. (\ref{fig:diag}) to the right. The Lagrangian of this model
is given in eq. \rf{lagr_4plet}. The calculation of the 
neutrino mass matrix for this model gives:
\begin{eqnarray}\label{eq:mnu3}
  (m_{\nu})_{\alpha\beta}&=& \frac{1}{16 \pi^2}
  \sum_j\sum_im_{\chi^{++}_j}
  (R_{S^{++}})_{1i} (R_{S^{++}})_{3i} (R_{\chi^{++}})_{1j} (R_{\chi^{++}})_{j2}^*
   \\ \nonumber
&\times &  B_0(0, m^2_{\chi_j^{++}}, m^2_{S^{++}_i}) \left[
(Y_1)_{\alpha}
(Y_2)_{\beta}
+
(Y_1)_{\beta}
(Y_2)_{\alpha}
\right]. \ \ \
\end{eqnarray}
Here $R_{S^{++}}$ and $R_{\chi^{++}}$ are the matrices which diagonalize the
doubly charged scalar and fermion mass matrices in the quadruplet model.
As in the triplet model, $\det(m_{\nu})=0$. Thus, the fit of neutrino
data is analogous to the one described above for the triplet model.
Recall, however, that in the numerical calculation we use 
$v_s$ as a free parameter.

\subsection{Lepton flavour violating decays}

As is well-known, experimental upper limits on lepton flavour
violating decays provide important constraints on TeV-scale extensions
of the standard model, see for example
\cite{Vicente:2015cka,Cai:2017jrq} and references therein. Flavour Kit
\cite{Porod:2014xia} implements a large number of observables into
SPheno \cite{Porod:2011nf}. In the following we will concentrate on
$\mu \to e \gamma$, $\mu \to 3 e$ and $\mu\to e$ conversion in Ti.

Currently $\mu \to e \gamma$ \cite{TheMEG:2016wtm} and $\mu\to 3 e$
\cite{Bellgardt:1987du} provide the most stringent constraints. There
is also a limit on muon conversion in Ti \cite{Dohmen:1993mp}.
However, while there will be only some improvement in the sensitivity
in $\mu \to e \gamma$ \cite{Baldini:2013ke}, proposals to improve
$\mu\to 3 e$ \cite{Blondel:2013ia} and muon conversion on both Ti
\cite{prime2003} and Al \cite{Pezzullo:2017iqq} exist, which claim
current bounds can be improved by 4-6 orders of magnitude. Constraints
involving $\tau$'s also exist, but are much weaker.  Thus, while we
routinely calculate constraints also for the $\tau$ sector, we will
not discuss the results in detail.

Again, let us first discuss the triplet model. The Lagrangian, see
eq. \rf{lagr_3plet}, of the model contains five different Yukawa
couplings. We can divide them into two groups: $Y_1$, $Y_2$ and $Y_3$
enter the neutrino mass calculation, while $Y_4$ and $Y_5$ are
parameters with no relation to $m_\nu$. This implies that for the
former, neutrino physics imposes a {\em lower bound} on certain
products of these Yukawas (as a function of the other parameters),
while the latter could, in principle, be arbitrarily small.

\begin{center}
\begin{figure}[ht]
\includegraphics[width=0.45\linewidth]{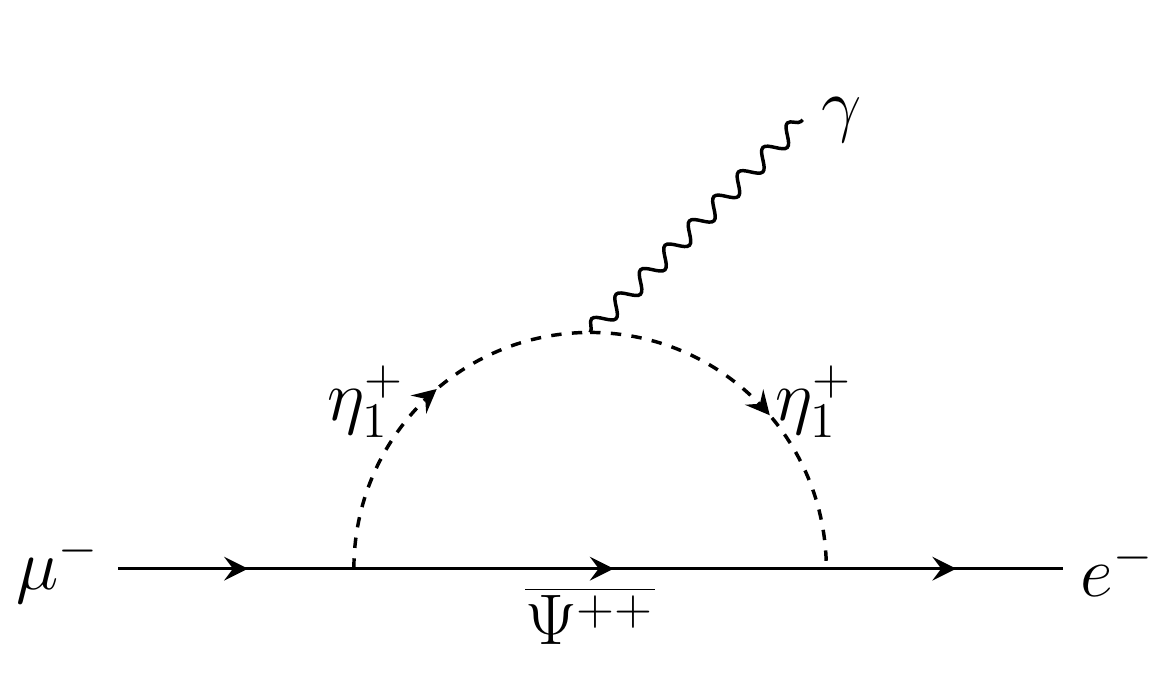}
\includegraphics[width=0.45\linewidth]{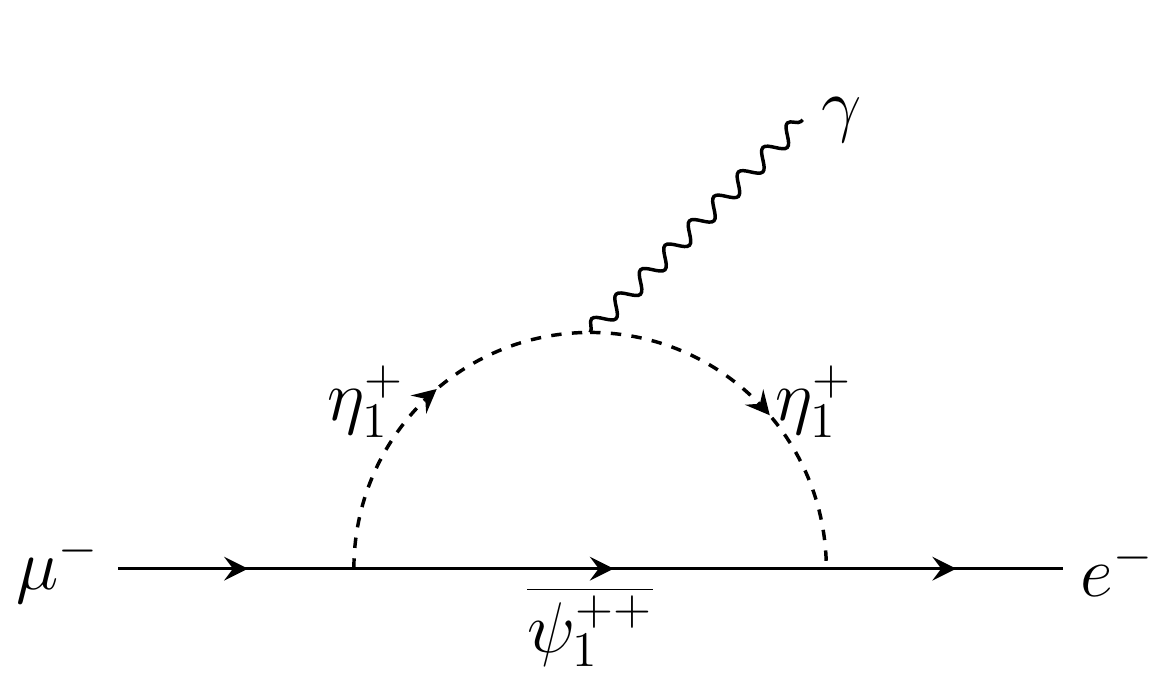}
\caption{Example diagrams for $\mu\to e \gamma$ in the triplet model,
  proportional to $(Y_{4})_e(Y_{4})_{\mu}$ (left) and
  $(Y_{5})_e(Y_{5})_{\mu}$ (right).}
\label{fig:diagmueg}
\end{figure}
\end{center}

Consider first the simpler case of $Y_4$ and $Y_5$. The diagrams in
fig. (\ref{fig:diagmueg}) show contributions to $\mu\to e \gamma$ due
to these couplings.  The current upper limit on Br($\mu \to e \gamma$)
then puts a bound on both, $Y_4$ and $Y_5$, of roughly $(Y_{4/5})_e
(Y_{4/5})_{\mu} \lsim 10^{-4}$ for masses of $\eta_1$ and $\Psi$ or
$\chi$ of the order ${\cal O}(1)$ TeV.

\begin{center}
\begin{figure}[t]
\includegraphics[width=0.4\linewidth]{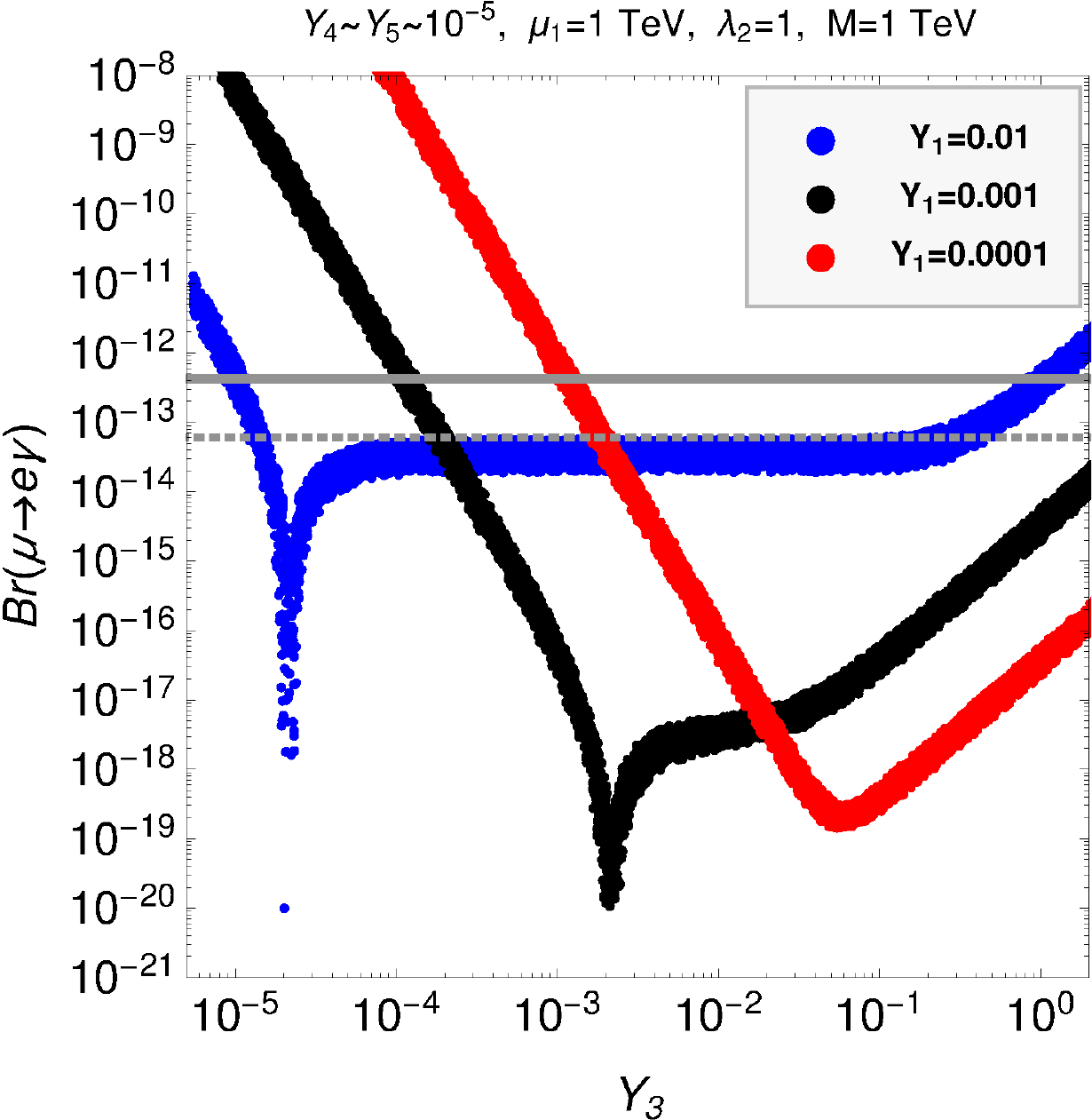}
\includegraphics[width=0.4\linewidth]{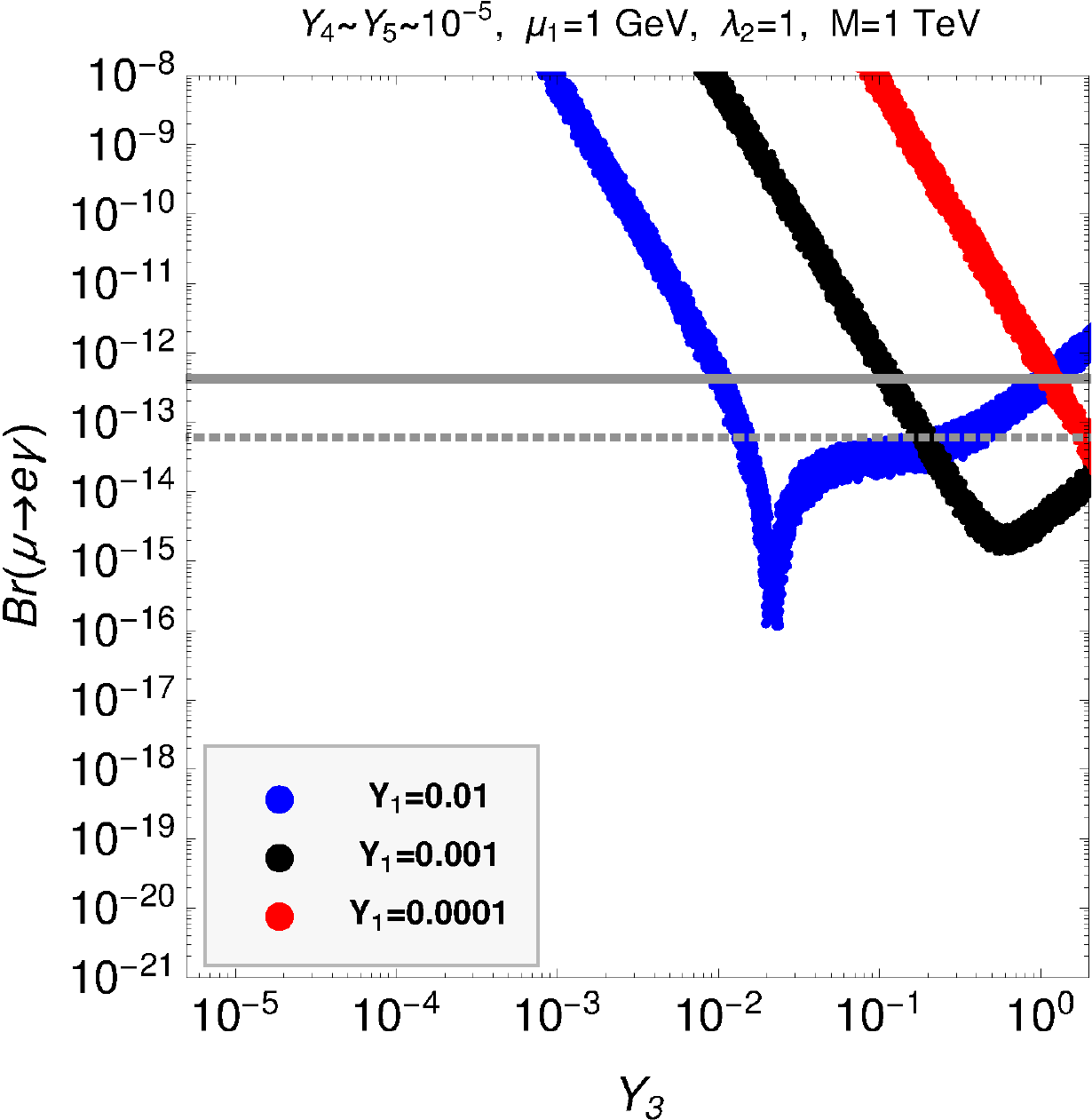}
\includegraphics[width=0.4\linewidth]{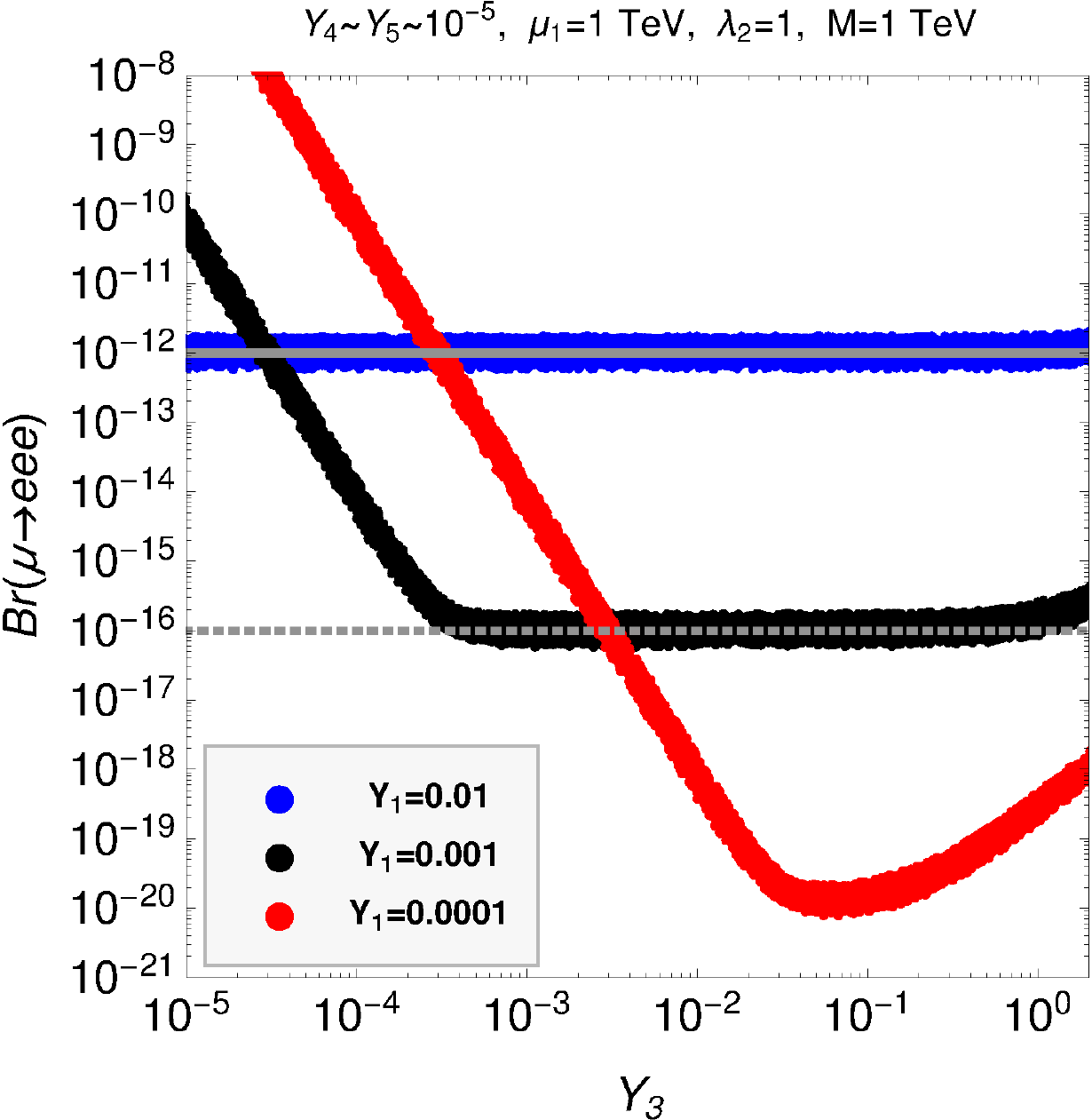}
\includegraphics[width=0.4\linewidth]{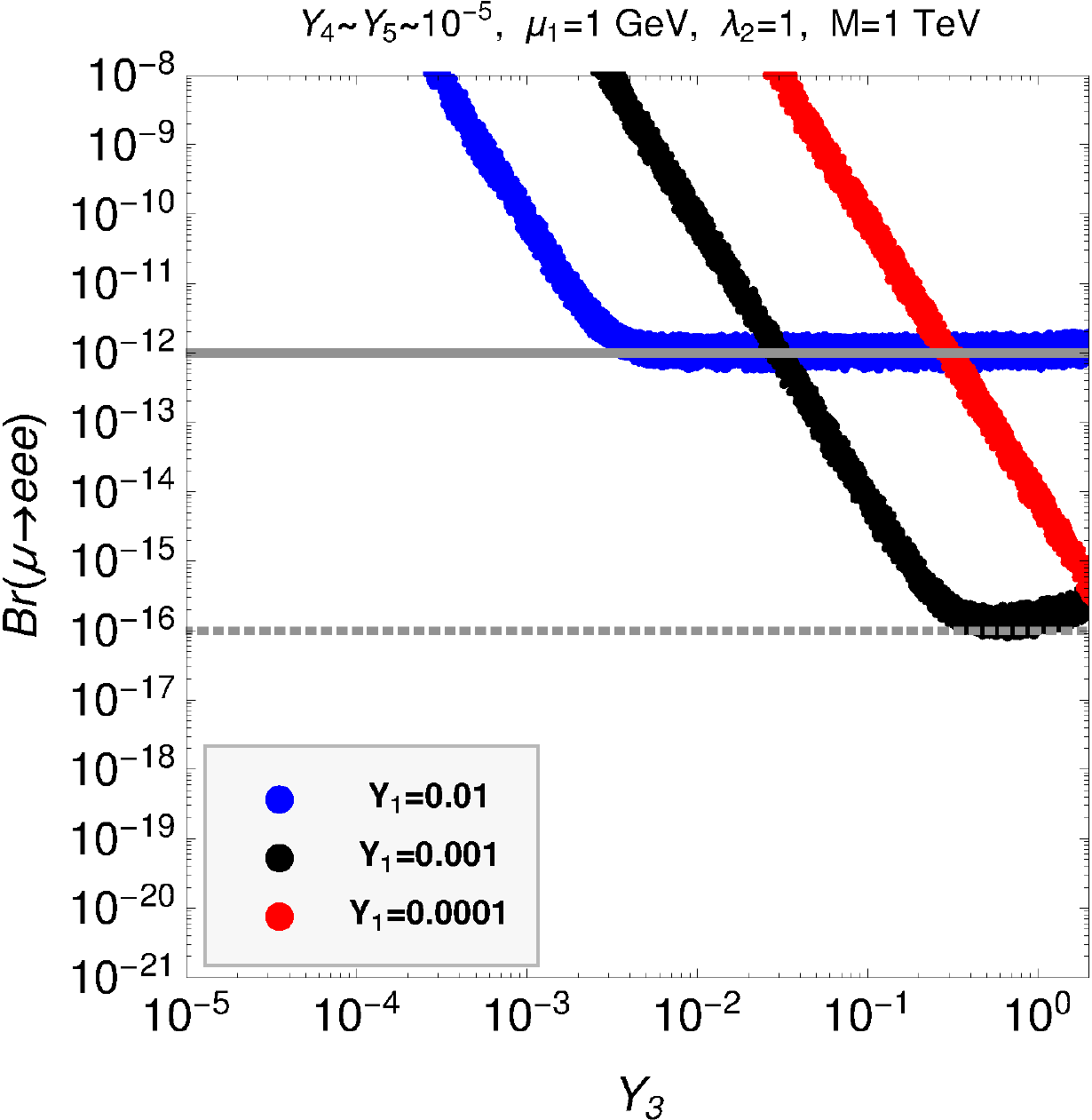}
\includegraphics[width=0.4\linewidth]{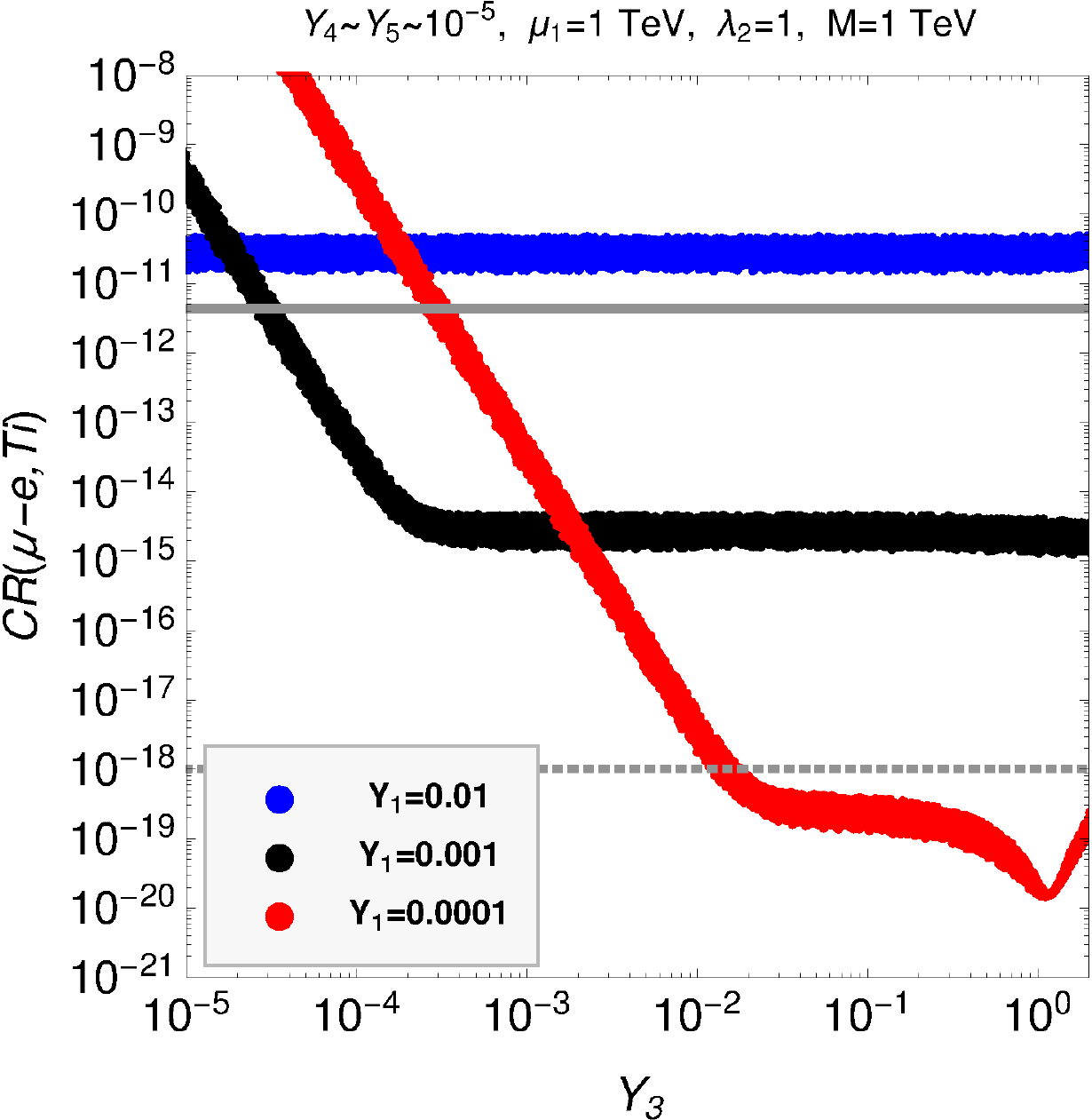}
\includegraphics[width=0.4\linewidth]{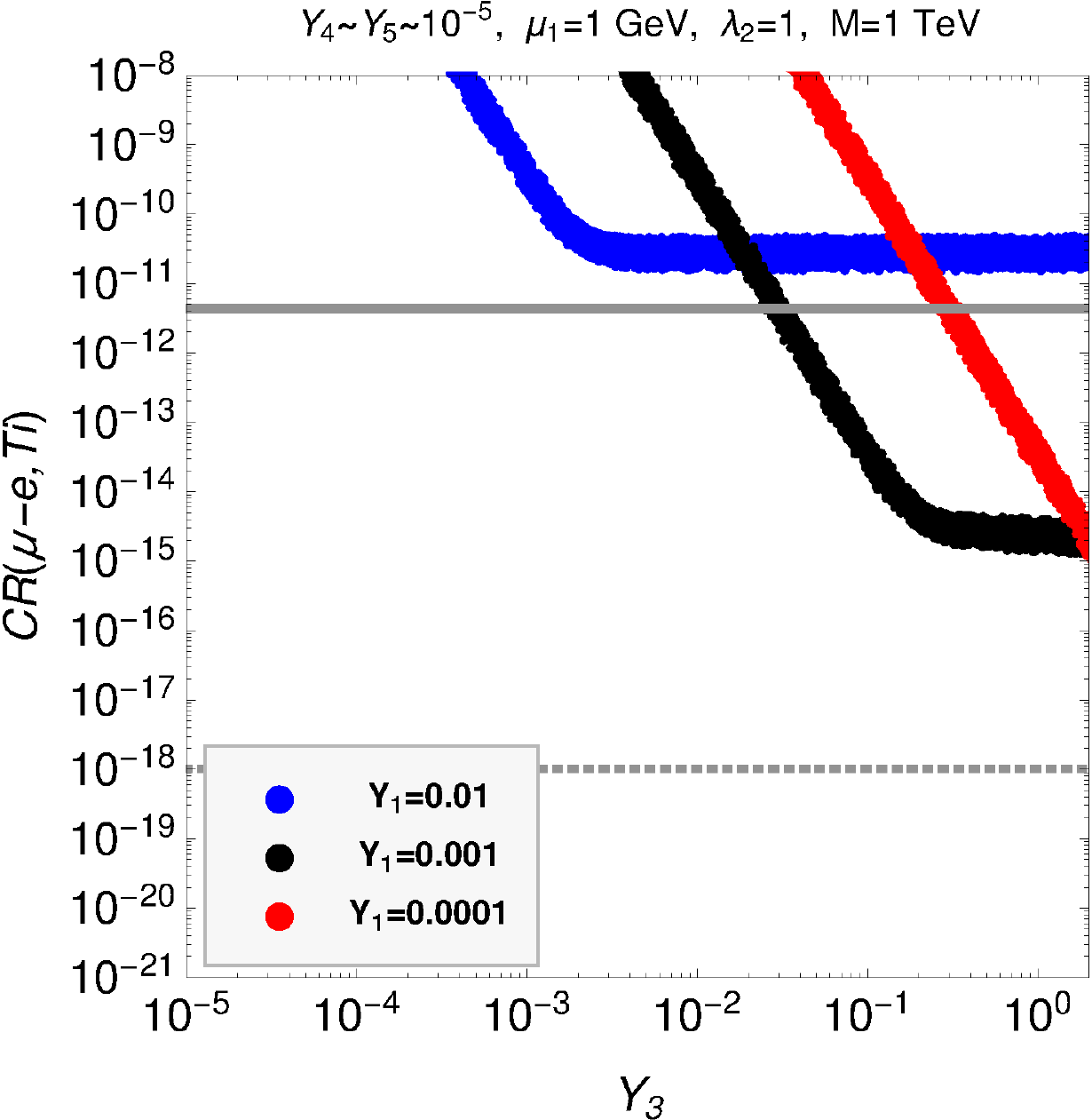}
\caption{Lepton flavour violating decays calculated in the triplet
  model.  Top panel: Br($\mu\to e \gamma$); middle panel: Br($\mu\to 3
  e$); bottom: $\mu\to e$ conversion in Ti. Rates are plotted versus the
  coupling $Y_3$, for discussion see text. Left row: $\mu_1=1$ TeV,
  right row $\mu_1=1$ GeV. The full (dashed) horizontal lines are the
  current limits (and future expected sensitivities).}
\label{fig:LFVT}
\end{figure}
\end{center}

The fit to neutrino data imposes relations among the parameters $Y_1$,
$Y_2$ and $Y_3$, see the discussion in the previous section. Thus, the
dependence of LFV decays on these parameters is slightly more
subtle. Fig. (\ref{fig:LFVT}) shows results for calculated branching
ratios of $\mu \to e \gamma$, $\mu\to 3 e$ and $\mu\to e$-conversion
in Ti, for several different choices of parameters, as function of
$Y_3$. The horizontal lines show current experimental limits (full
lines) and future expected sensitivities (dashed lines). Note that
$Y_3$ has no lepton flavour indices and, thus, by itself can not
generate a LFV diagram. Instead, for fixed values of masses and the
parameters $\lambda_2$ and $\mu_1$, the prefactor ${\cal F}$
determining the size of the calculated neutrino masses, see
eq. (\ref{eq:prefac}), depends linearly on $Y_3$. Keeping neutrino
masses constant while varying $Y_3$, thus leads to a corresponding
change in (the inverse of) $Y_1 \times Y_2$. For this reason, for
small values of $Y_3$ the branching ratios in fig.  (\ref{fig:LFVT})
decrease with increasing $Y_3$. For the largest values of $Y_3$,
diagrams with additional $Y_3v/m_{\Psi}$ insertions can become
important and branching ratios start to rise again as a function of
$Y_3$. Note that in all calculations in fig. (\ref{fig:LFVT}), we have
chosen $Y_4$ and $Y_5$ small enough, such that their contribution to
the LFV decays is negligible.

Both, $Y_1$ and $Y_2$, generate LFV decays.  Whether diagrams
proportional to $(Y_1)_e(Y_1)_{\mu}$ or to $(Y_2)_e(Y_2)_{\mu}$ give
the more important contribution to $\mu\to e \gamma$ depends on the
(mostly) arbitrary choice of $(Y_1)_e$. In fig. (\ref{fig:LFVT}) we
plot results for three different choices of $(Y_1)_e$. For $(Y_1)_e =
10^{-2}$ there is a large range of $Y_3$, for which $\mu\to e \gamma$
and $\mu\to 3 e$ remain constant. In this case, diagrams proportional
to $(Y_1)_e(Y_1)_{\mu}$ dominate the partial width.

We also show in fig. (\ref{fig:LFVT}) two different choices of the
parameter $\mu_1$. To the left: $\mu_1 = 1$ TeV, to the right $\mu_1
=1$ GeV. Smaller values of $\mu_1$ require again larger values of the
Yukawa coupling $Y_2$, and thus lead to larger LFV decays. While for
$\mu_1 = 1$ TeV nearly all points in the parameter space are allowed
with current constraints, once $(Y_1)_e$ is smaller than roughly (few)
$10^{-3}$, for $\mu_1 =1$ GeV large parts of the parameter space are
already ruled out. For $\mu_1 \simeq 10^{-2}$ GeV and masses below 2
TeV there remain already now no valid points in the parameter space
which, at the same time, can obey upper limits from $\mu \to e \gamma$
and explain neutrino masses, except in the small regions where different
diagrams cancel each other exactly accidentally.

It is worth to mention that for the triplet model the branching ratio
of $\mu \rightarrow 3e$ is higher than the corresponding of $\mu
\rightarrow e\gamma$. Naively one would expect the former to be two
orders (an order of $\alpha_{EM}$) lower than the latter. However,
$\mu \rightarrow e\gamma$ occurs at loop level, while in this model
there exists a tree level diagram for $\mu \rightarrow 3e$ mediated by
a $Z^0$, due to the mixing between leptons and $\overline{\Psi^+}$, so
proportional to $(Y_1)_e (Y_1)_{\mu}$. Other tree level contributions
mediated by doubly-charged scalars are also possible due to this mixing. 
These are proportional to $(Y_4)_{\mu} (Y_4)_e (Y_1)_e (Y_1)_e$, so the 
upper limit given by $\mu \rightarrow e \gamma$ is still dominant.

The plots in fig. (\ref{fig:LFVT}) also show the discovery potential
of future $\mu\to 3e$ and $\mu$-conversion experiments. In particular,
an upper bound on $\mu$ conversion of the order $10^{-18}$ would
require both, very small Yukawas (for example: $(Y_1)_e \lsim
10^{-5}$) and a large value of $\mu_1 \gsim 1$ TeV at the same
time. All other points in the parameter space of the triplet model
(assuming they explain neutrino data) with masses below 2 TeV, should
lead to the discovery of $\mu$-conversion. This is an interesting
constraint, since such small values of the Yukawa couplings would
imply very long lived particles at the LHC. We will come back to this
discussion in the next section.

\begin{center}
\begin{figure}[t]
\includegraphics[width=0.4\linewidth]{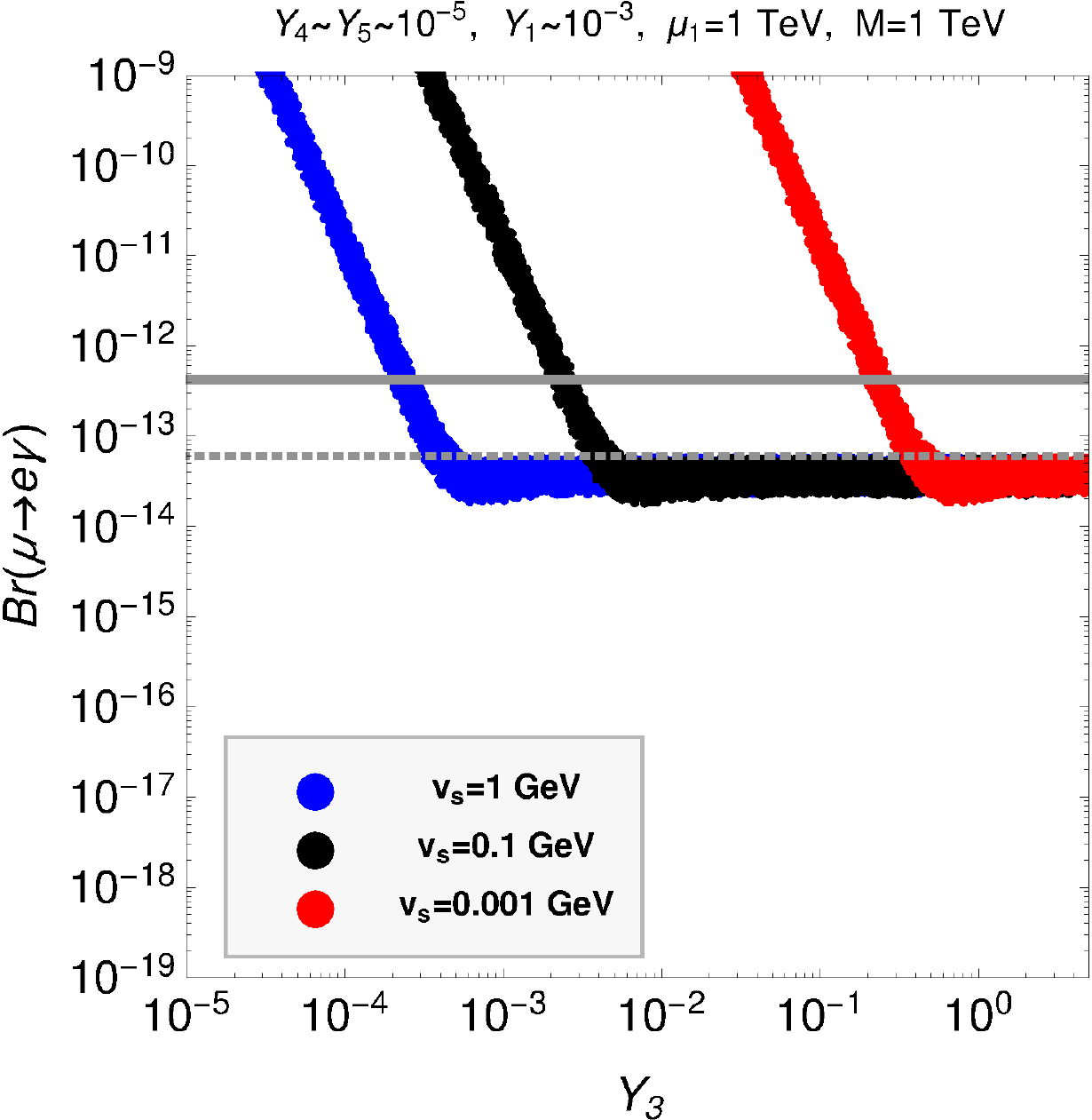}
\includegraphics[width=0.4\linewidth]{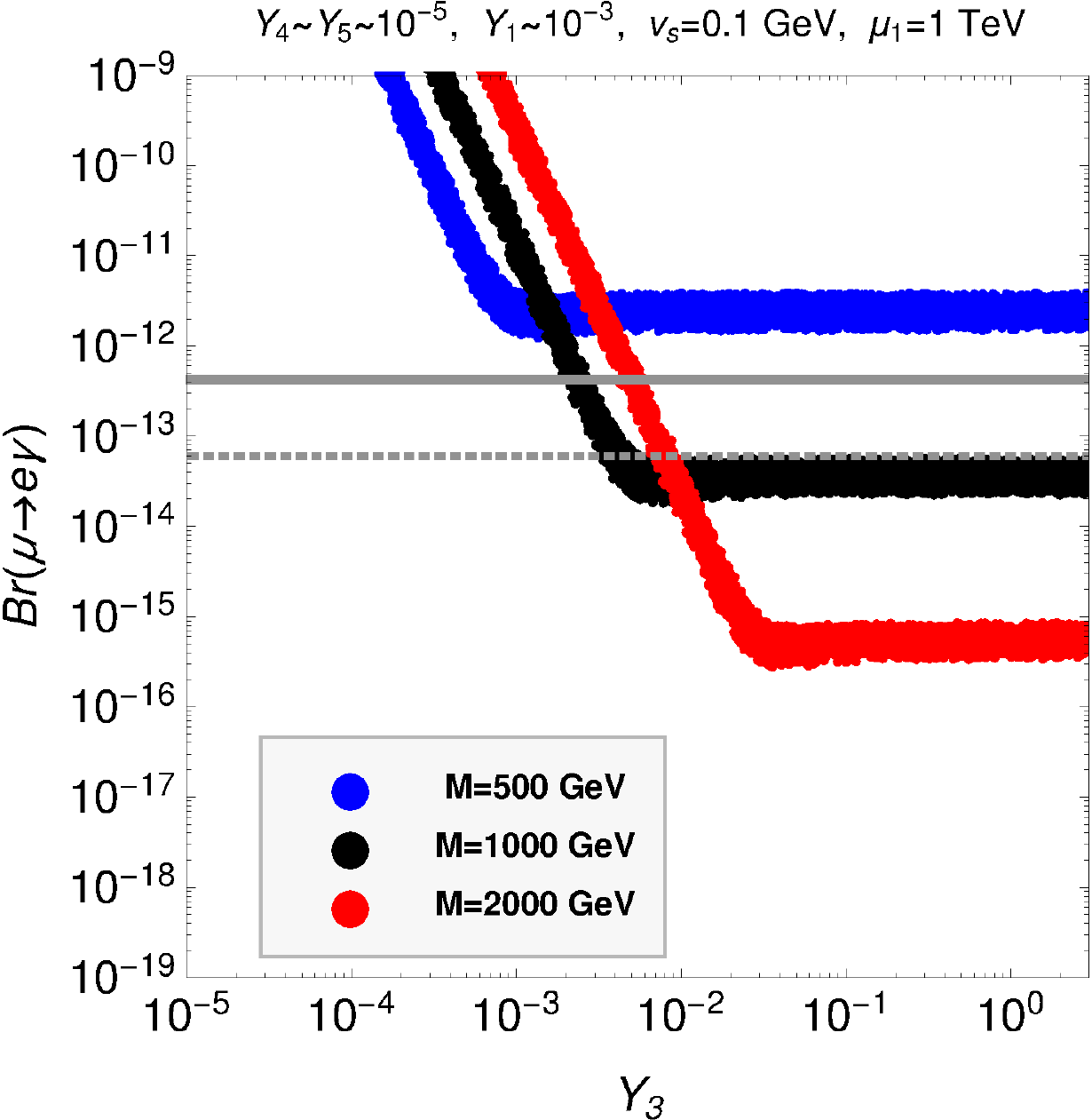}
\includegraphics[width=0.4\linewidth]{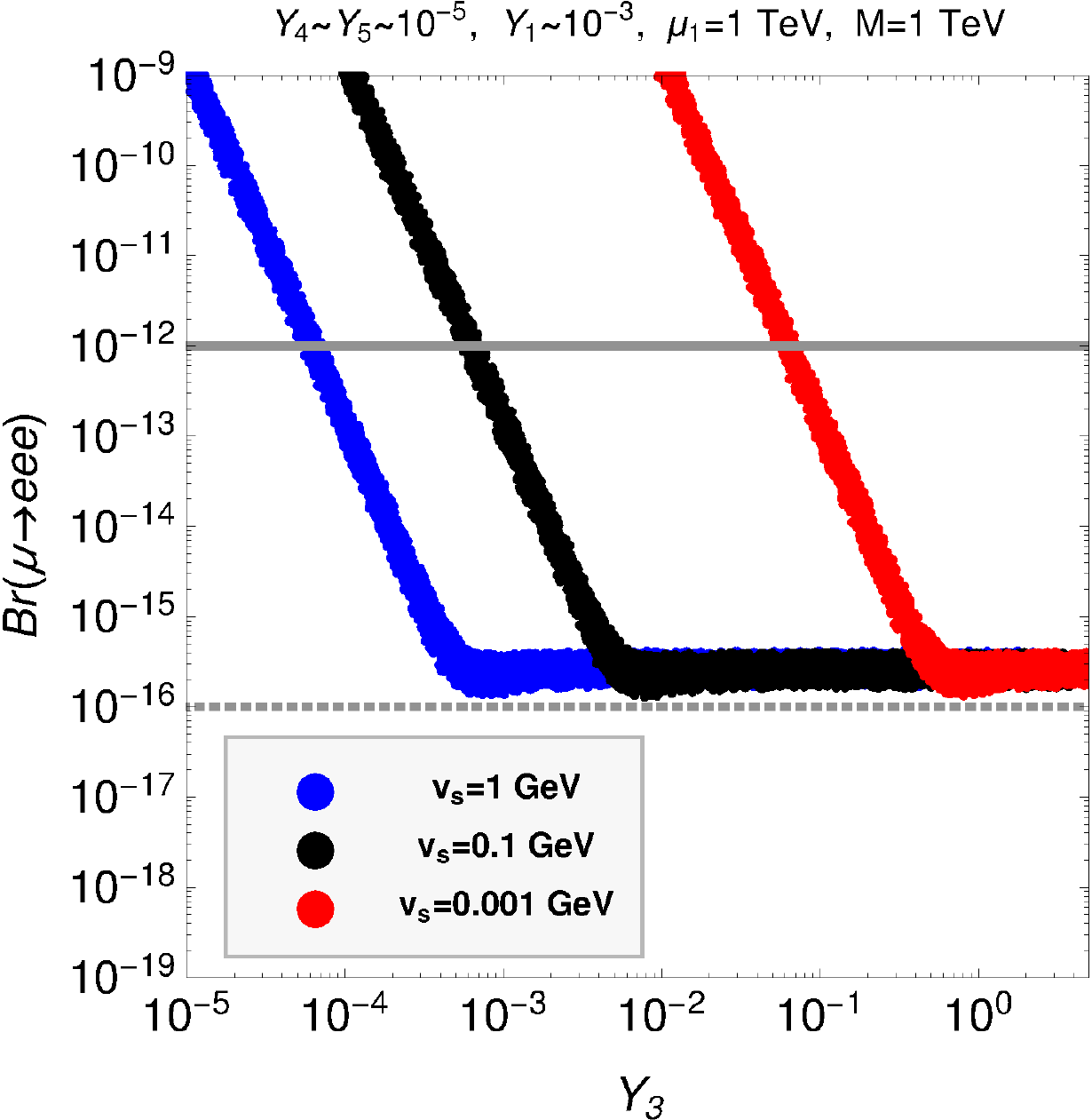}
\includegraphics[width=0.4\linewidth]{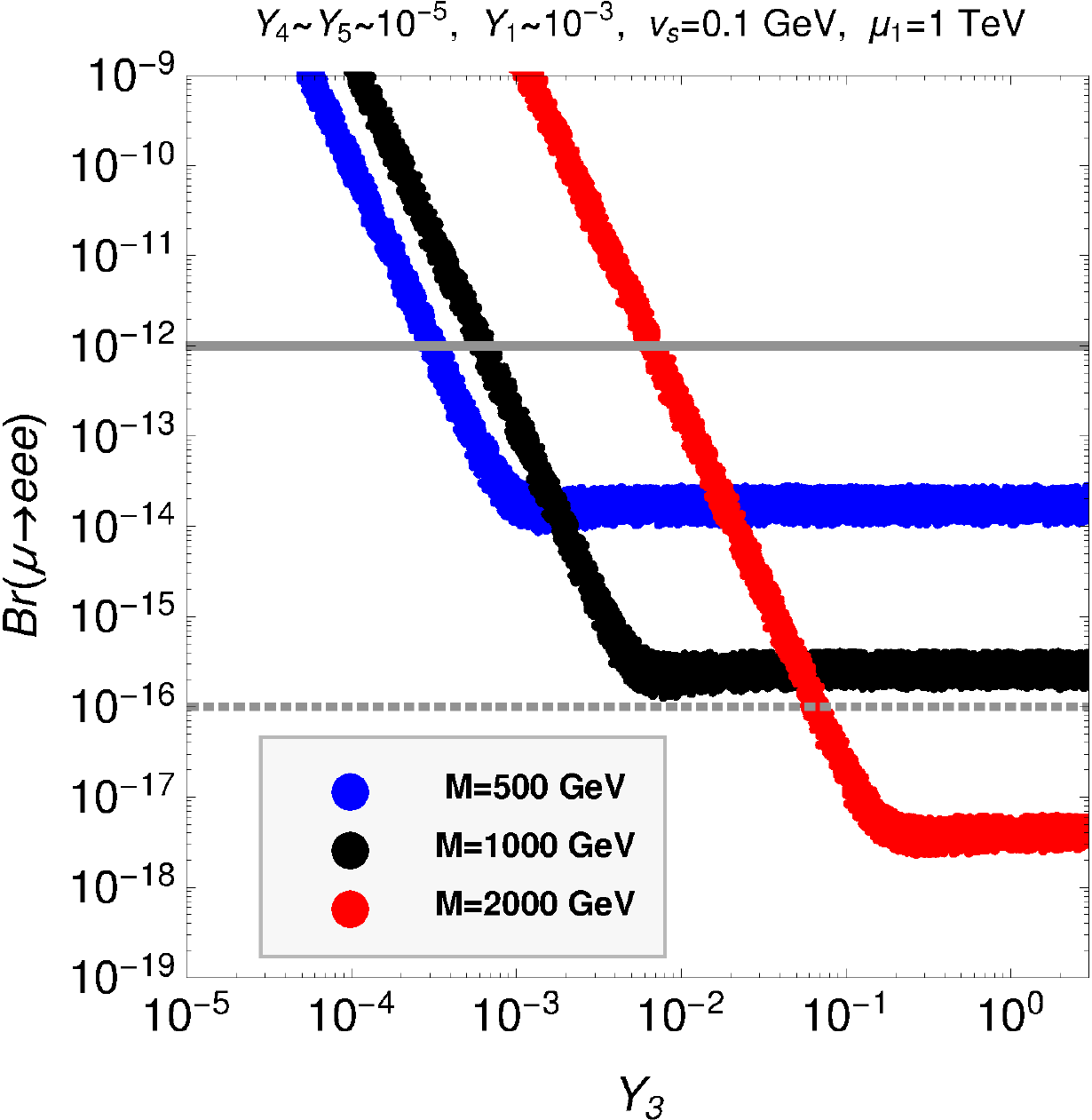}
\includegraphics[width=0.4\linewidth]{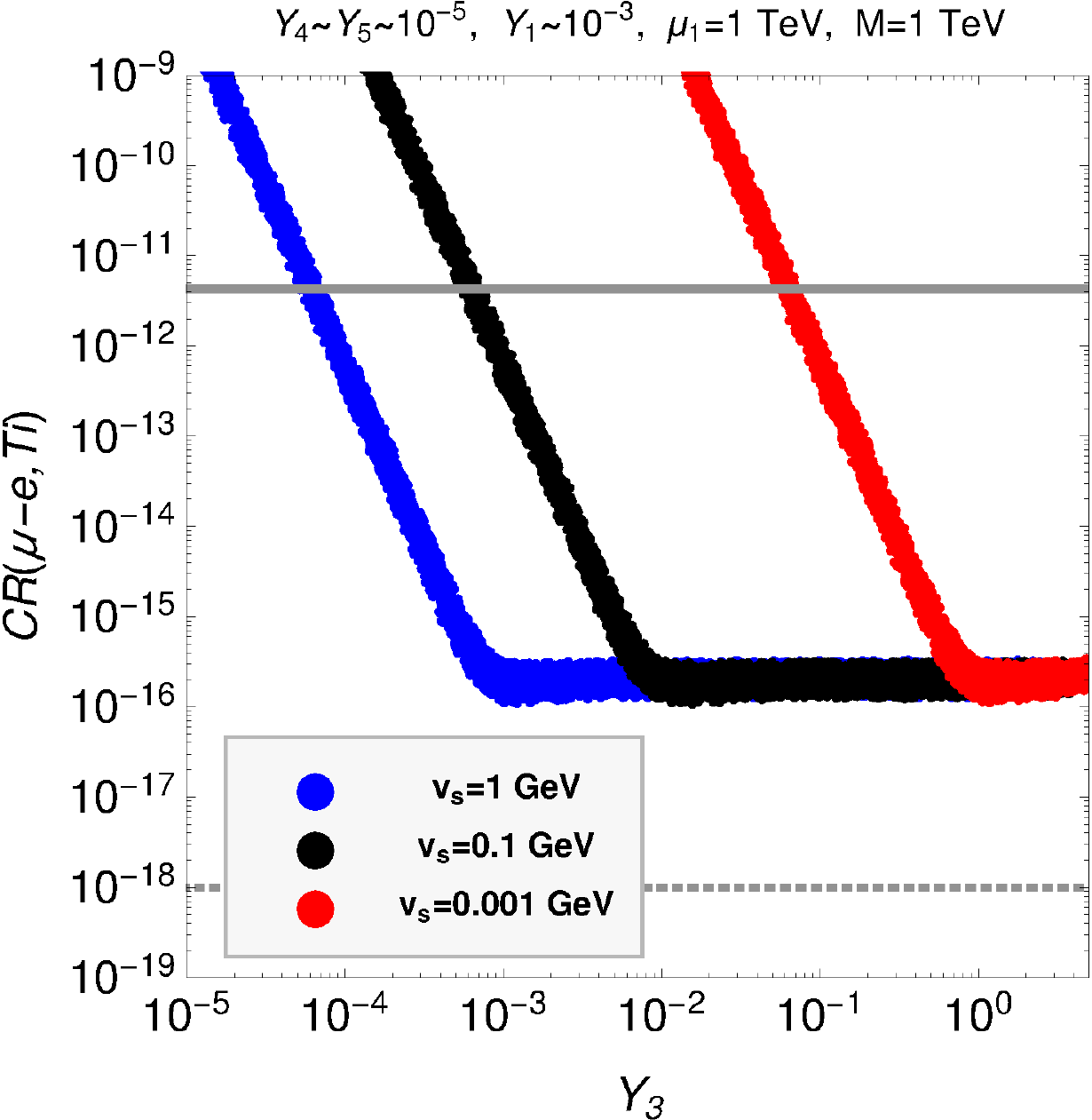}
\includegraphics[width=0.4\linewidth]{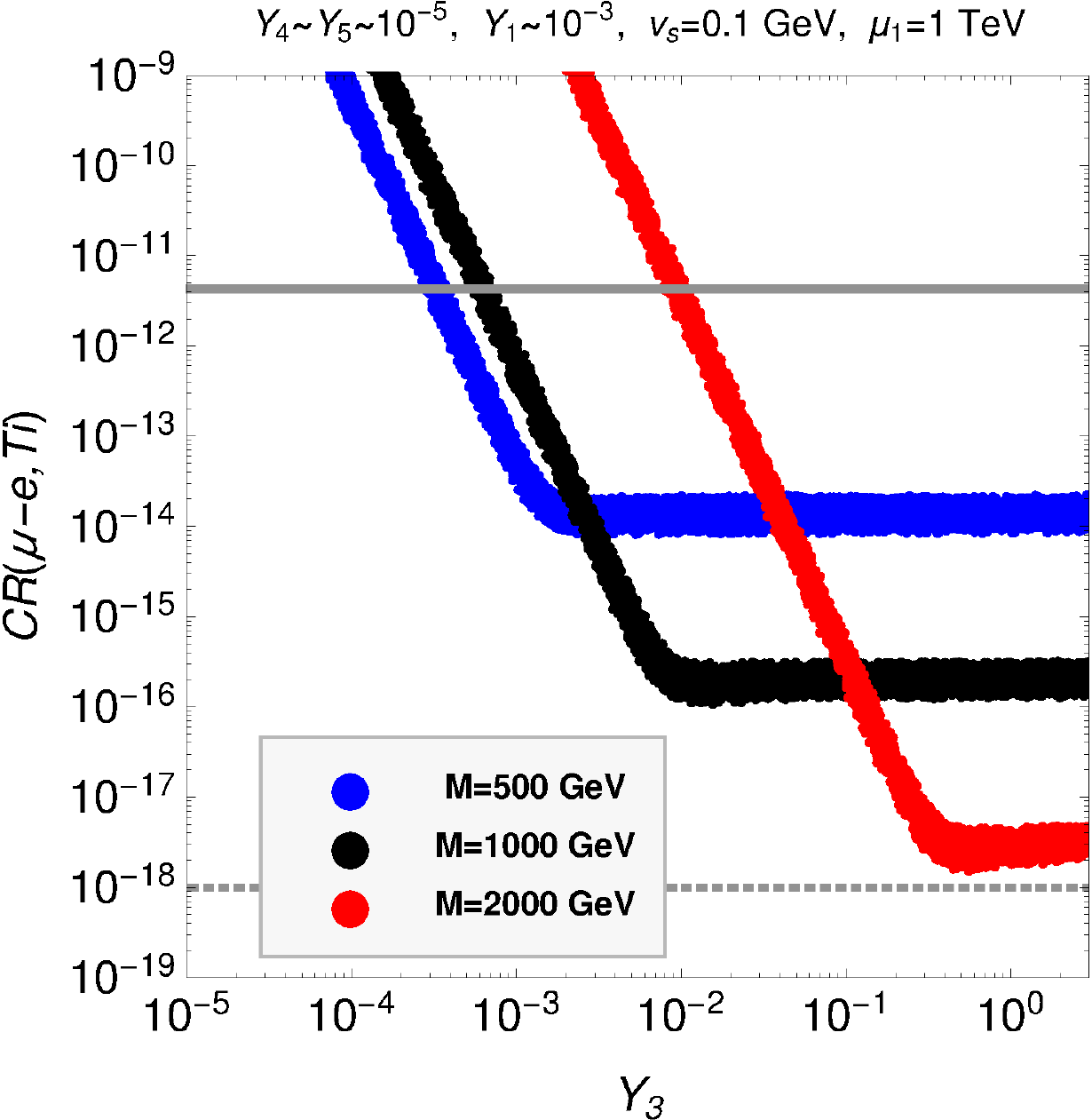}
\caption{Lepton flavour violating decays calculated in the quadruplet
  model. Plots to the left show results for different choices of the
  quadruplet vev $v_S$, while the ones to the right use different
  masses for the new scalars and fermions. Here, we assume that all
  new particles have similar masses of the order indicated in the
  figure panels.}
\label{fig:LFVQ}
\end{figure}
\end{center}

We now turn to a discussion of LFV in the quadruplet model. Similarly to
the triplet model, we can divide parameters into two groups: $Y_1$-$Y_3$
depend on the neutrino mass fit, while $Y_4$-$Y_6$ are unconstrained
parameters. Constraints on $Y_4$ and $Y_5$ from LFV are very similar
to those found in the triplet model. The constraints on $Y_6$ are
somewhat more stringent, $(Y_6)_{\mu e}(Y_6)_{ee} \lsim 10^{-5}$, since
there exists a tree-level diagram via doubly-charged scalar exchange
contributing to the decay $\mu \to 3 e$.

Turning to $Y_1$-$Y_3$, fig. (\ref{fig:LFVQ}) shows some sample
calculations of LFV decays as function of $Y_3$ in the quadruplet
model. The plots to the left show $\mu \to e \gamma$, $\mu\to 3 e$ and
$\mu$-conversion in Ti, for several different choices of the
quadruplet vev $v_S$. Smaller values of $v_S$ need larger values of
the Yukawa couplings $Y_2$ for constant neutrino masses. Thus, LFV
decays are larger at the same values of $Y_3$ for smaller values of
$v_S$.
The plots on the right of fig. (\ref{fig:LFVQ}) show the same LFV
decays, for a fixed value of $v_S=0.1$ GeV, but different values of
the new scalar and fermion masses.  As simplification in this plot we
assume that all new scalars and fermions have roughly the same mass,
$M$, as indicated in the plot panels.  Larger values of masses lead to
smaller LFV decay widths, as expected. As also is the case for the
triplet model, future bounds from $\mu\to 3 e$ and $\mu\to e$-conversion
will test most of the relevant parameter space of the quadruplet model
up to masses of order 2 TeV. 

In fact, even for masses as large as 2 TeV, non-observation of $\mu\to
e$ conversion would put an interesting lower limit on the value of
$v_S$, which we roughly estimate to be around $v_S=0.1$ GeV.  Note
that there is an upper limit on $v_S$ from the SM $\rho$ parameter of
the order of $v_S \lsim 2.5$ GeV \cite{Babu:2009aq}.

In summary, the non-observation of LFV decays can be used to put
upper bounds on the Yukawa couplings of our models. At the same time
the observed neutrino masses require lower bounds on these Yukawa
couplings and the combination of both constraints result in a very
restricted range of allowed parameters. We have shown this explicitly
only for our two example models, but the same should be true for
any of the possible (genuine) $d=7$ 1-loop models.

\section{Phenomenology at the LHC\label{Setc:LNV}}

\subsection{Constraints from LHC searches}

We have calculated the production cross sections for the different
scalars and fermions of our example models using MadGraph
\cite{Alwall:2007st,Alwall:2011uj}. Pair production is usually
calculated via s-channel photon and $Z^0$ exchange, while associated
production, such as $\eta^{--}\eta^{+++}$, proceeds via $W^+$
diagrams. However, as pointed out in \cite{Ghosh:2017jbw}, for large
masses the pair production cross section of charged particles via
photon-photon fusion can give the dominant contribution to the cross
section, despite the small photon density in the proton. In our
calculation we use the NNPDF23$\_$nlo$\_$as$\_$0119 parton distribution
function, which contains NLO corrections, necessary for inclusion of
the photon-photon fusion contributions.  We have checked numerically
and find that at the largest masses cross sections can be enhanced up
to one order of magnitude for multiply charged particles. For this
reason we concentrate on pair production of particles in the
following. Note, however, that for lower masses (up to roughly 1 TeV),
associated production is large enough to produce additional signals,
not discussed here.

Results for the cross sections are shown in fig.  (\ref{fig:Prod}) for
$\sqrt{s}=13$ TeV. To the left we show results for scalars, to the
right the cross sections for fermions. The scalar cross sections (to
the left) where calculated for the scalars of the triplet model.  The
fermion cross section (to the right) correspond to the fermions of the
quadruplet model.  The underlying Lagrangian parameters were chosen
such, that the corresponding gauge states (index shown in the figure)
are the lightest mass eigenstate of the corresponding charge. Cross
sections do also depend, to some extent, on the hypercharge of the
particle. However, since photon-fusion dominates the cross section at
large values of the masses, all mass eigenstates with the same
electric charge have similar cross sections. We therefore do not
repeat those plots for all particles in our models.

For the quadruply charged particles of the models cross sections
larger than $10^{-2}$ fb are obtained, even for masses up to 2.5 TeV.
Note that at the largest value of masses pair production cross section
ratios for differently charged particles simply scale as the ratio
of the charges to the 4th power. We will come back to this in the
discussion of the LNV signals in the next subsection. 

\begin{center}
\begin{figure}[tbph]
\begin{centering}
\includegraphics[scale=0.6]{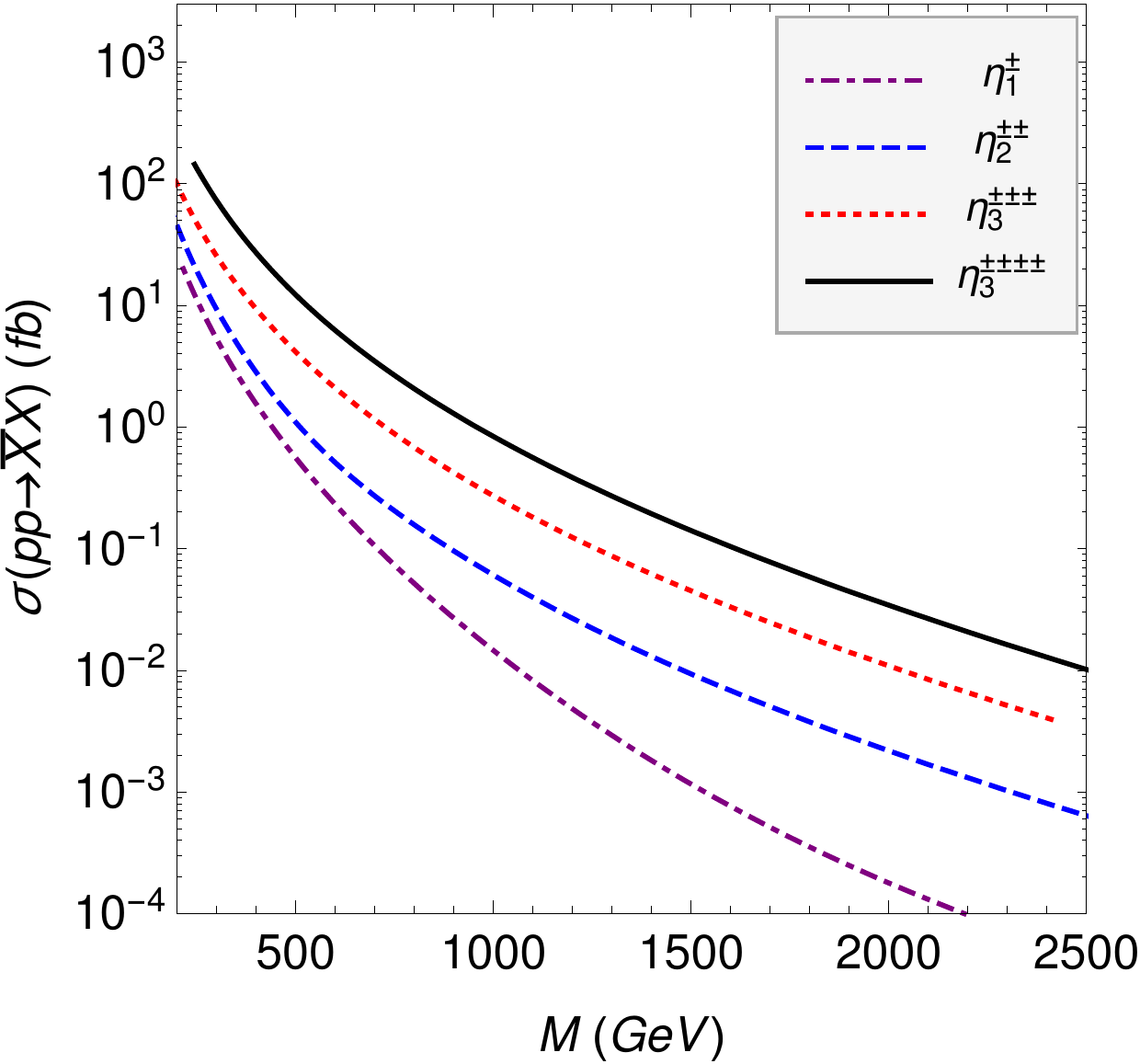}
\includegraphics[scale=0.6]{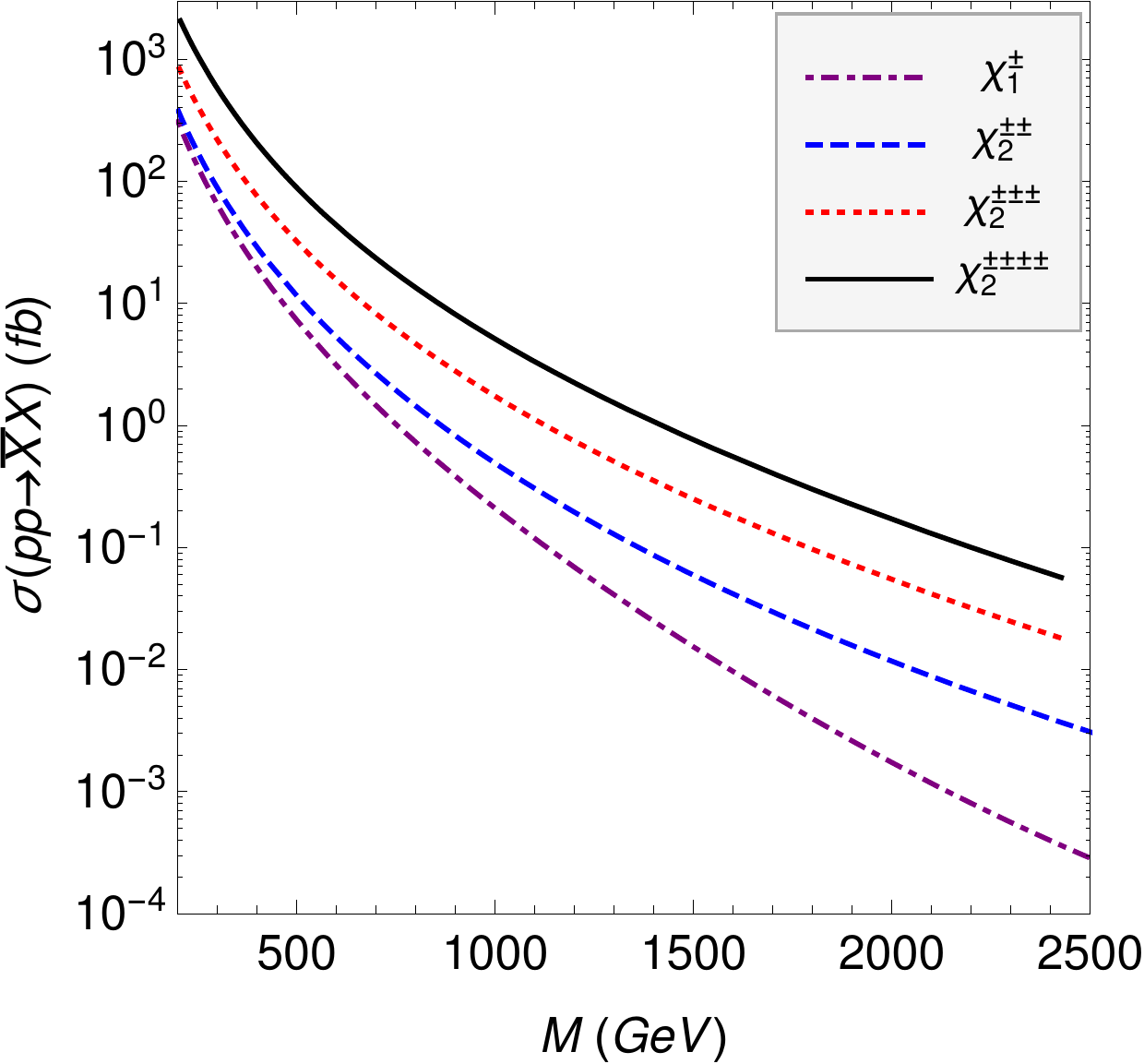}
\end{centering}
\protect\caption{\label{fig:Prod} Pair production cross sections for the
  different scalars (left) and fermions (right) of the two example models.
For discussion see text.}
\end{figure}
\par
\end{center}

A number of different LHC searches can be used to set limits on the
various particles of our example models. The simplest search, and
currently the most stringent LHC limit for our models, comes from a
recent ATLAS search for doubly charged particles decaying to either
$e^{\pm}e^{\pm}$, $e^{\pm}\mu^{\pm}$ or $\mu^{\pm}\mu^{\pm}$ final
states \cite{ATLAS:2017iqw}.  Results of our calculation, compared to
the experimental limit are shown in fig. (\ref{fig:Limpp}) for the
$\mu^{\pm}\mu^{\pm}$ final state.

The two-body decay with of the doubly charged scalar $\eta_1^{++}$ is
approximately given by:
\begin{equation}\label{eq:GamPP}
  \Gamma(\eta_1^{++}\to l^{+}_{\alpha}l^{+}_{\beta}) \simeq
  \frac{1}{8\pi} \left( \frac{v}{m_{\Psi}} \right)^2
  \left[ (Y_4)_{\alpha}(Y_1)_{\beta}+(Y_4)_{\beta}(Y_1)_{\alpha} \right]^2 m_{\eta^{++}_1}
\end{equation}
Since the Yukawa coupling $Y_4$ does not enter the neutrino mass
calculation, the exact value and flavour composition of this decay can
not be predicted. However, $Y_1$ enters our neutrino mass fit.  The
observed large neutrino angles require that all entries in the vector
$Y_1$ are different from zero and of similar order. Typically, from
the fit we find numerically ratios in the range
$(Y_1)_e:(Y_1)_{\mu}:(Y_1)_{\tau} \sim ([1/4,1/2]:[1,3]:1)$, but the
exact ratios depend on the allowed range of neutrino angles. Scanning
over the allowed neutrino parameters then leads to a variation of the
branching ratios of the $\eta_1^{++}$ into the different lepton
generations.  This explains the spread of the numerically calculated
points in fig. (\ref{fig:Limpp}). Combined with the experimental limit
from ATLAS, lower mass limits in the range of (600-800) GeV
result. Note that in this plot, we allow all three neutrino angles to
float within the 3 $\sigma$ regions of the global fit
\cite{Forero:2014bxa}.

\begin{center}
\begin{figure}[tbph]
\begin{centering}
  \includegraphics[scale=0.6]{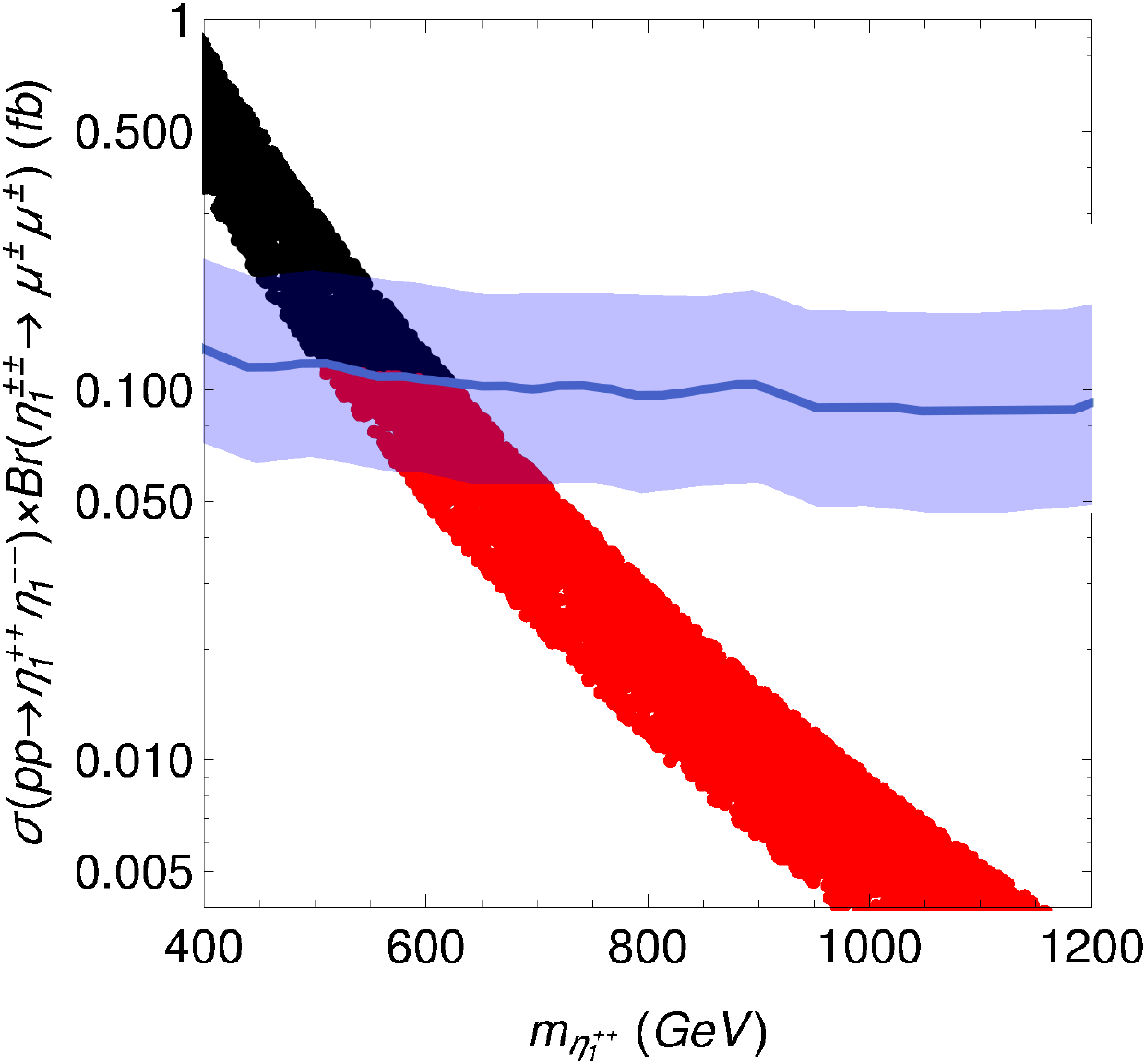}
  \end{centering}
\protect\caption{\label{fig:Limpp} Current constraints on doubly
  charged scalars, using the recent search by ATLAS
  \cite{ATLAS:2017iqw}.  The blue line is the limit quoted in
  \cite{ATLAS:2017iqw}, the light blue region the 95 \%
  c.l. region. Points are our calculation, scanning over the allowed
  ranges of neutrino angles. Red points are allowed by this search.}
\end{figure}
\par
\end{center}

\begin{center}
\begin{figure}[tbph]
\begin{centering}
  \includegraphics[scale=0.4]{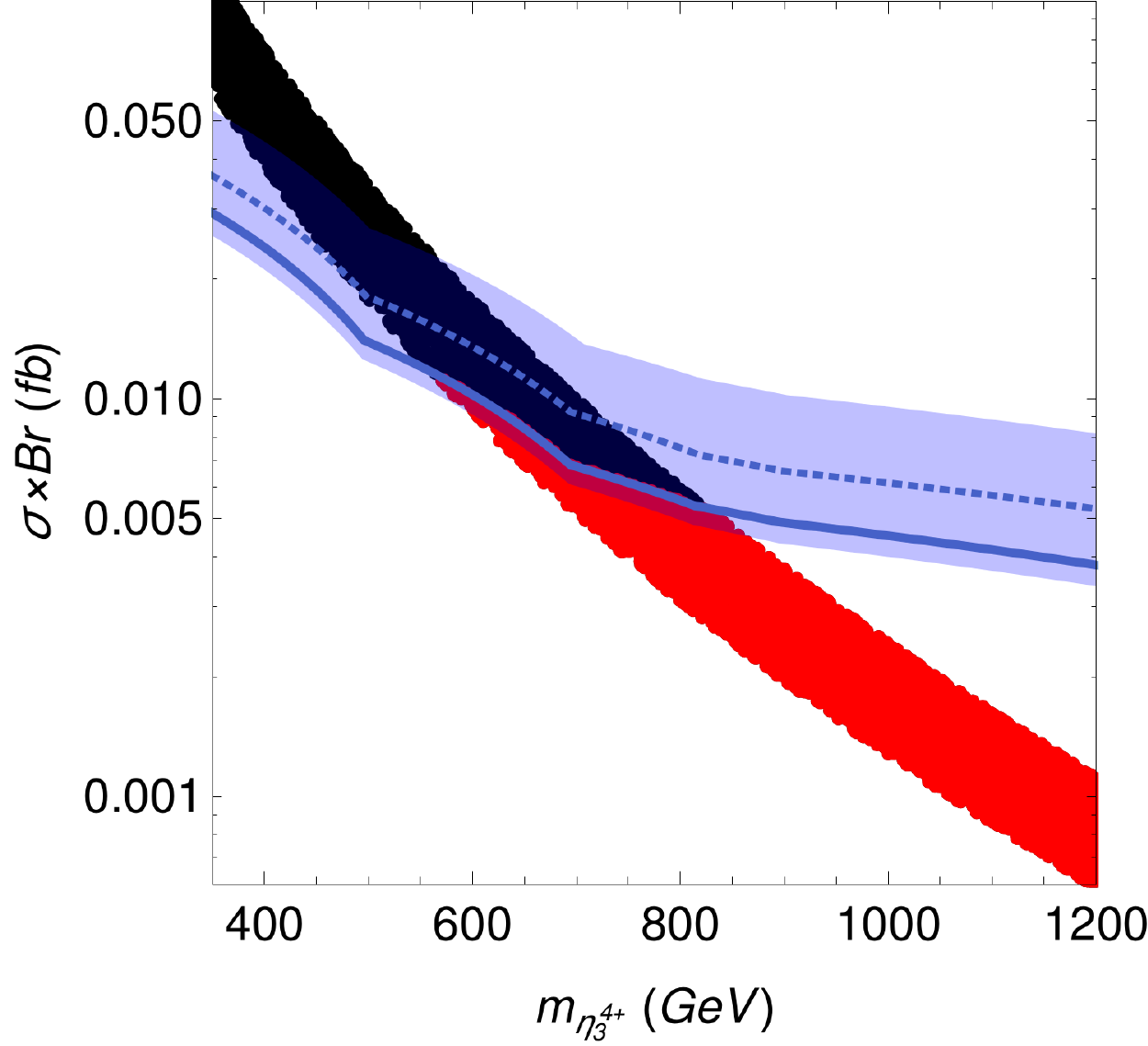}
  \includegraphics[scale=0.4]{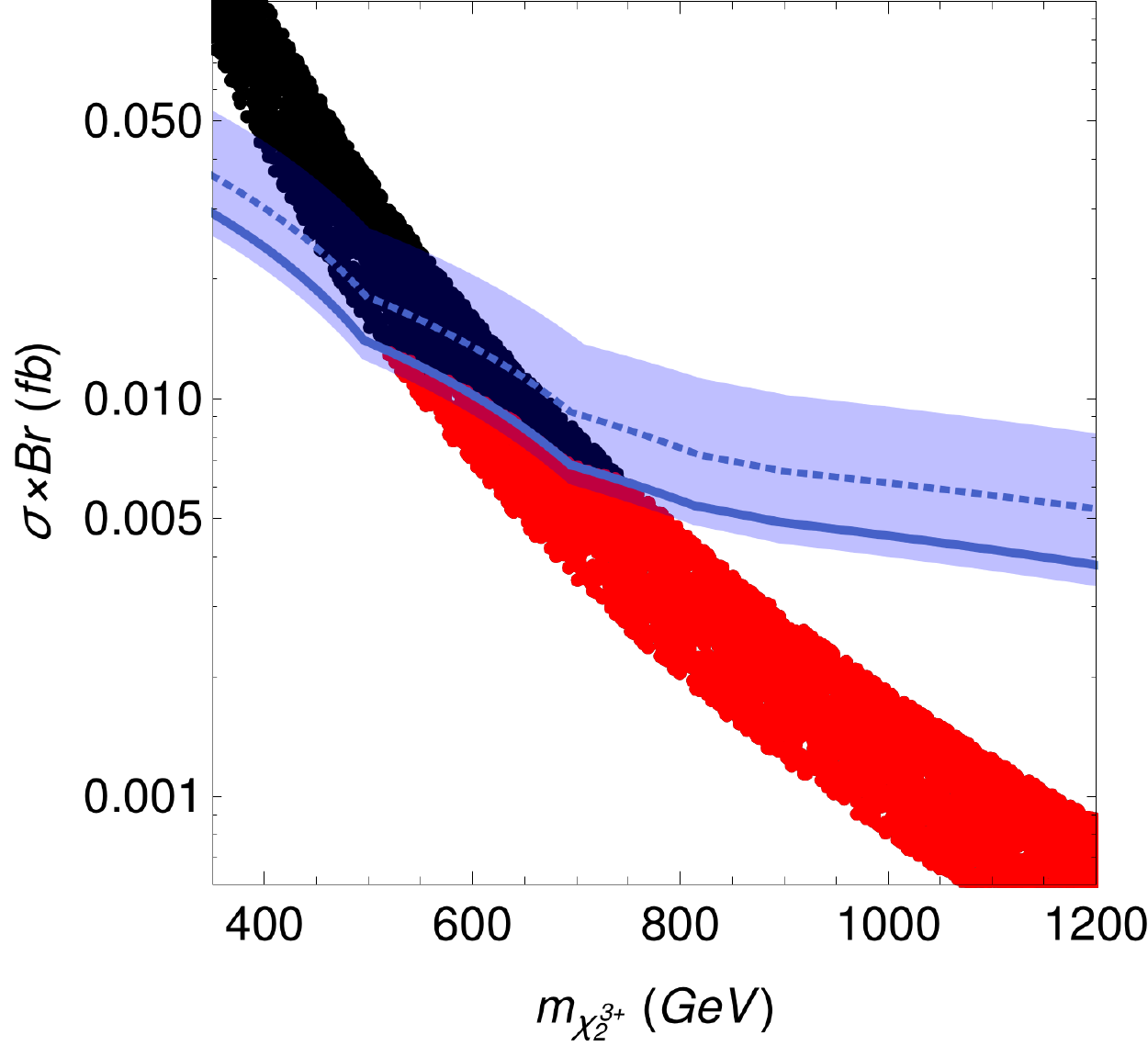}
  \includegraphics[scale=0.4]{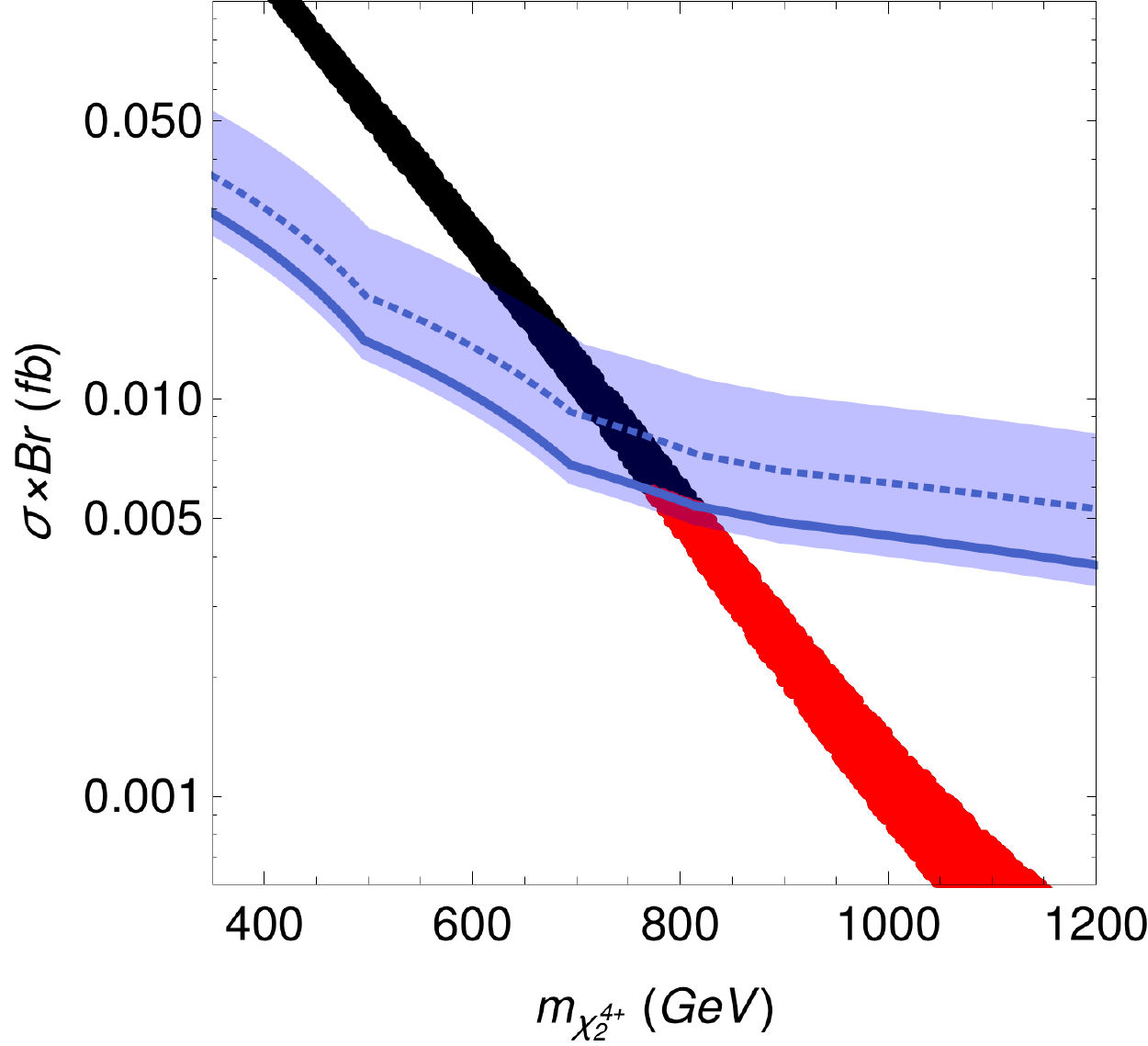}
  \end{centering}
\protect\caption{\label{fig:MultiLep} Current constraints on charged
  scalars and fermions using the multi-lepton search
  \cite{Sirunyan:2017qkz}. Points are our numerical results, the bands are
  experimental limits, see also fig. (\ref{fig:Limpp}).}
\end{figure}
\par
\end{center}

The CMS collaboration has recently published a search based on
multi-lepton final states \cite{Sirunyan:2017qkz}. The original motivation
for this search is the expectation that the fermions of the seesaw
type-III lead to final states containing multiple charged leptons and
missing momentum.  For example, $\Sigma^{\pm}\Sigma^0 \to W^{\pm}\nu
W^{\pm}l^{\mp}$ from the associated production of the fermionic
triplet $\Sigma=(\Sigma^{+},\Sigma^0,\Sigma^{-})$. The analysis
\cite{Sirunyan:2017qkz} requires than at least three charged leptons plus
missing energy and takes into account both, electrons and muons.

In our models, these final states can occur in various decay chains.
Consider for example $\chi_2^{4+}$. Once produced, it can decay into a
$\chi_1^{3+}+W^+$, which further decays to a doubly charged
scalar and $l^+$. The doubly charged scalar decays to either leptons
or $W$'s. The missing energy is then produced in the leptonic decays
of the $W$'s. Here, all intermediate particles can be either on-shell
or off-shell, depending on the unknown mass hierarchies. Constraints
can then be derived from the results of \cite{Sirunyan:2017qkz}, scanning
over the allowed ranges of the branching ratios, which lead to a least
three charged leptons plus at least one $W$ in the final state. 

In fig. (\ref{fig:MultiLep}) we show results of this procedure for
the examples of $\eta_3^{4+}$, $\chi_2^{3+}$ and $\chi_2^{4+}$. The
lower limits, derived from this exercise, have a rather large
uncertainty, due to the unknown branching ratios. For example,
the lower mass limit for $\eta_3^{4+}$ is in the range of (550-850)
GeV. Note that $\eta_3^{4+}$ could decay, in principle to four
charged leptons with a branching ratio close to 100 \%. The final
state from pair production of $\eta_3^{4+}$ would then contain
eight charged leptons and missing momentum would appear only from the
decays of the $\tau$'s. In this case, our simple-minded recasting
of the multi-lepton search \cite{Sirunyan:2017qkz} ceases to be valid
and the lower limit on the mass of $\eta_3^{4+}$, mentioned above
does not apply. As fig. (\ref{fig:MultiLep}) shows, the lower limit
on the mass of $\chi_2^{4+}$ is more stringent than the one for
$\eta_3^{4+}$. This simply reflects the larger production cross sections
for fermions, compare to fig. (\ref{fig:Prod}).

\subsection{New LNV searches}

\begin{table}[h]  
\begin{center}
  \begin{tabular}{|c|c|c|c|c|c||c|c|c|c|}
    \hline
Multiplicity  &LNV Signal & Particles& Model & Mass range  \\
\hline
4 & $l^{\pm} l^{\pm}  +  W^{\mp} W^{\mp}$ & $S^{{\pm} {\pm}}$, $\phi_{1}^{{\pm}{\pm}}$, $\phi_{2}^{\pm\pm}$& Q & $m < 1.4$ TeV  \\
\hline
  6 & $l^{\pm} l^{\pm} W^{\pm} +  W^{\mp} W^{\mp} W^{\mp}$ & $S^{3+}$, $\phi_{2}^{3+}$& Q &   $m < 2.0$  TeV\\
\hline
6 & $l^{\pm} l^{\pm} l^{\pm}  +   W^{\mp} W^{\mp} l^{\mp}$ & $\chi_{2}^{3+}$& Q & $m < 2.6$  TeV \\
\hline
8 & $l^{\pm} l^{\pm} W^{\pm} W^{\pm}+   W^{\mp} W^{\mp} W^{\mp} W^{\mp}$ & - & -&  \\
\hline
8 & $l^{\pm} W^{\pm} W^{\pm} W^{\pm}+   l^{\mp} l^{\mp} l^{\mp} W^{\mp}$ & $\chi_2^{4+}$ & Q & $m < 3.2$ TeV\\
\hline
 8 & $l^{\pm} l^{\pm} l^{\pm} l^{\pm}+  l^{\mp} l^{\mp} W^{\mp} W^{\mp}$ & $\eta_{3}^{4+}$& T & $m < 2.5$ TeV \\
\hline
\hline
\end{tabular}
\end{center}
\caption{\it List of ``symmetric'' LNV final states in $d=7$
  models. The first column counts the number of final state particles,
  the second column gives the LNV signal. Here, we have separated the
  total final state into the two sets of particles, coming from the
  pair produced states listed in column 3. The invariant masses of the
  quoted subsystems should peak at the mass of the particle quoted in
  column 3. Column four gives the model in which this signal could be
  found. The last column gives our simple estimate for the mass range,
  which can be probed at the LHC with ${\cal L}\simeq 300$/fb. For a
  discussion see text.
\label{Tablelnv}}
\end{table}

We now turn to a discussion of possible LNV signals at the LHC. Table
\ref{Tablelnv} shows examples of different LNV final states from pair
production of scalars or fermions in the two models under
consideration. This list is not complete since (a) associated
producion of particles is not considered; (b) the table gives only
``symmetric'' LNV states, see below, and (c) we do not give LNV final
states with neutrinos, since such states do not allow to establish LNV
experimentally. 

The table gives in column 1 the multiplicity of the final state and in
column 2 the LNV signal. In that column, the two possible final states
from the decay of the particle given in column 3 are given seperately.
The invariant masses of both separate subsystems in column 2, should
therefore give peaks in the mass of the particle in column 3.

Particles in column 3 are quoted as gauge eigenstates. However,
scalars in our models are, in general, admixtures of different gauge
eigenstates.  Consider, for example, the simplest final state
$l^{\pm}l^{\pm} +W^{\mp}W^{\mp}$. $\phi_1^{\pm\pm}$ can decay to
$l^{\pm}l^{\pm}$, via the coupling $Y_6$, while $S^{\pm\pm}$ can decay
to $W^{\pm}W^{\pm}$ via the induced vev $v_S$ (or, equivalently
proportional to $\lambda_2$). The doubly charged scalars mix via the
entries in the mass matrices proportional to $\mu_1$, $\lambda_3$ (and
$\lambda_4$), see eq. \rf{pot_4plet}. Whether the lightest doubly charged
mass eigenstate is mostly $\phi_1$, $\phi_2$ or $S$ depends on the
choice of parameters, but the results are qualitatively very similar
in all cases. We therefore show in fig. (\ref{fig:LNV}) only
the results for the case where $S_1^{++}$ is mostly $S$. 

\begin{center}
\begin{figure}[tbph]
\begin{centering}
  \includegraphics[scale=0.5]{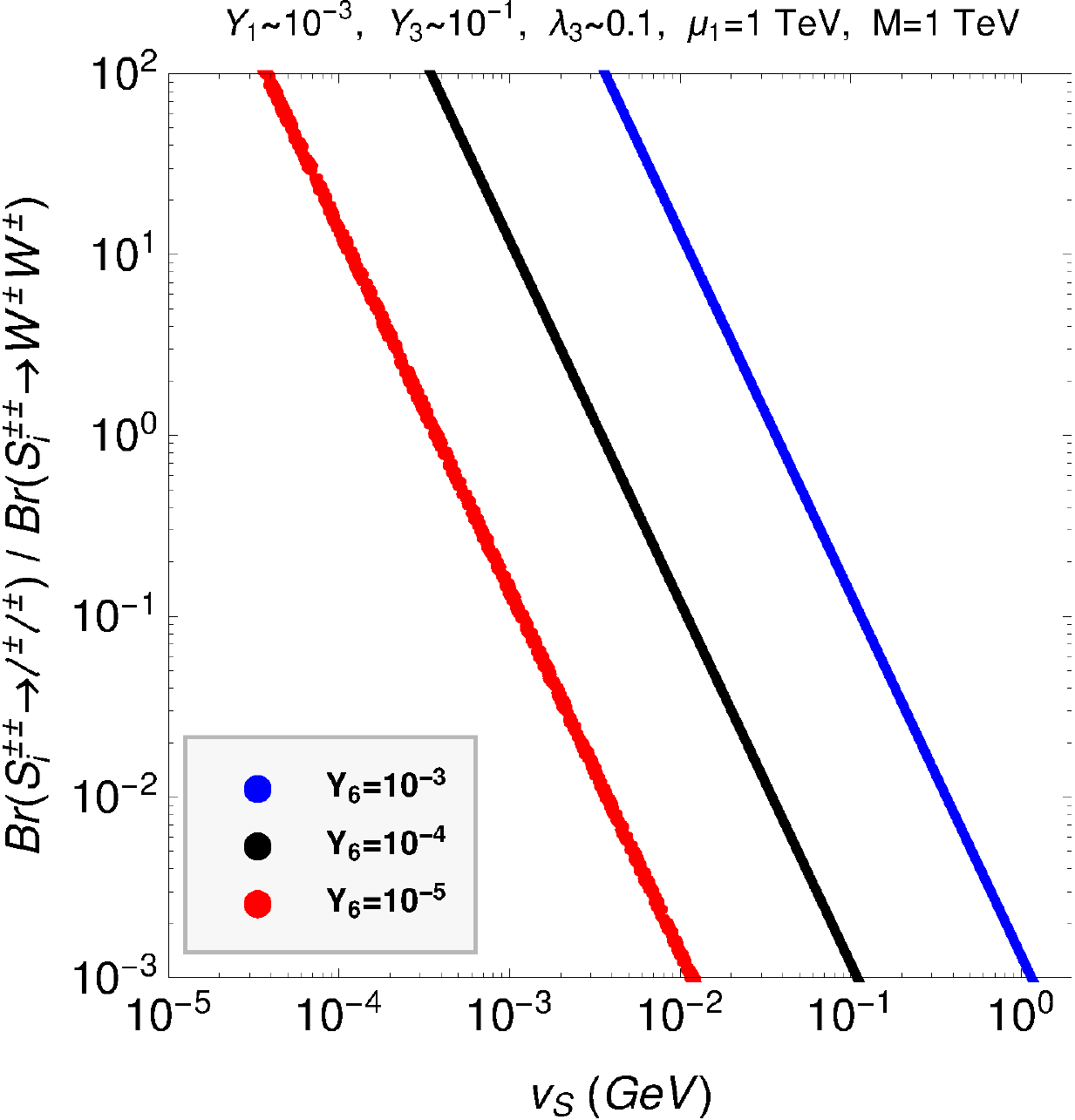}\hskip3mm
  \includegraphics[scale=0.5]{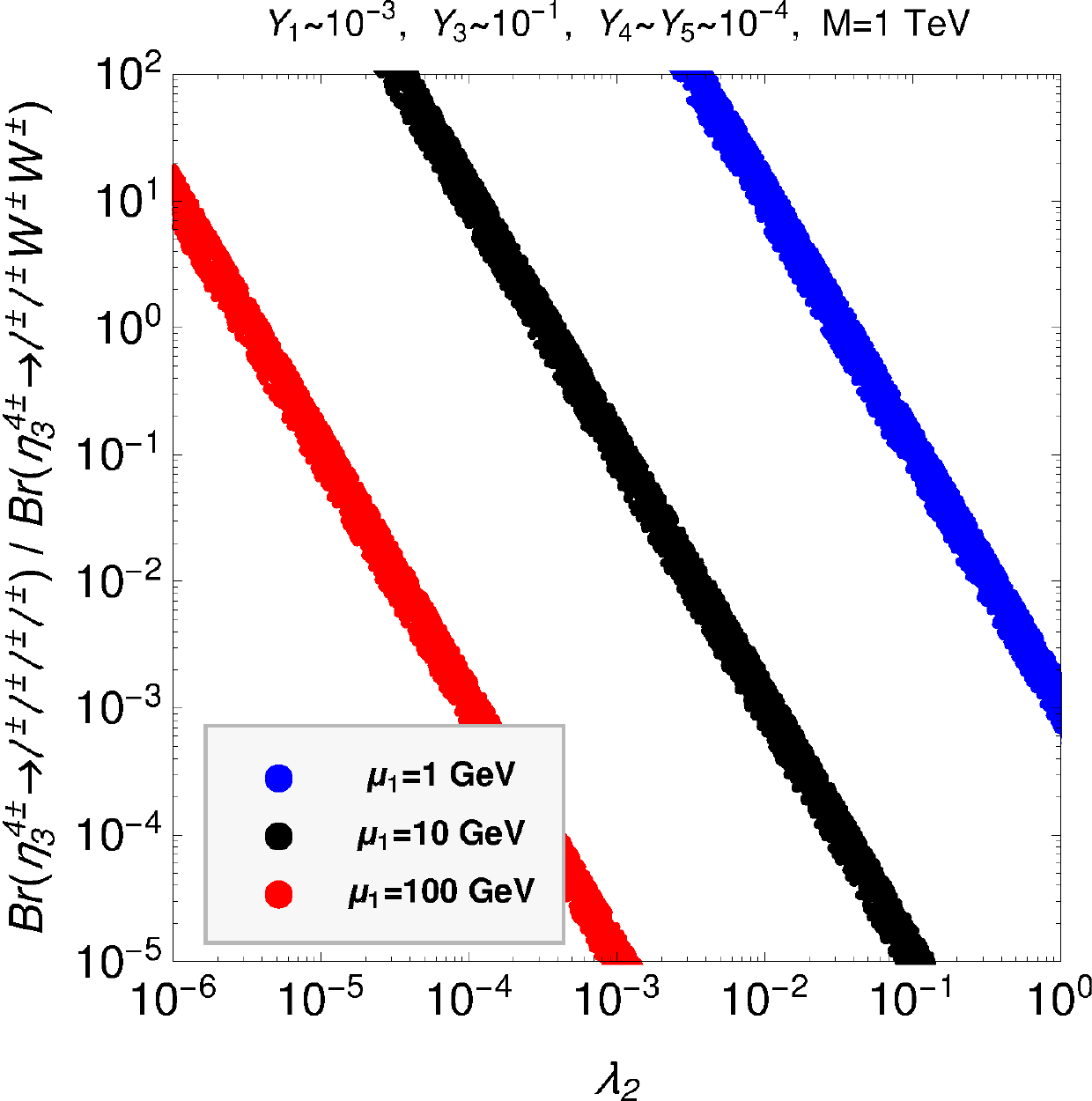}
  \end{centering}
\protect\caption{\label{fig:LNV} To the left: Ratio of branching
  ratios of the doubly charged scalar, $S_1^{++}$ decaying to
  $l^{\pm}l^{\pm}$ divided by $W^{\pm}W^{\pm}$ as a function of $v_S$
  for some fixed choice of the other model parameters and three
  different values of $Y_6$. This plot assumes that the lightest
  doubly charged scalar $S_1^{++}$ is mostly the gauge state
  $S^{++}$. Results for the other cases are qualitatively very similar
  and thus not repeated.  To the right: Ratio of branching ratios for
  $\eta_3^{4+}$ decays. As in the case of $S_1^{++}$, LNV will be
  observable only if this ratio is of order ${\cal O}(1)$.}
\end{figure}
\par
\end{center}

Fig. (\ref{fig:LNV}) (left) shows the ratio of branching ratios of the doubly
charged scalar, $S_1^{++}$ decaying to $l^{\pm}l^{\pm}$ divided by the
decay to $W^{\pm}W^{\pm}$ as a function of $v_S$ for some fixed choice
of the other model parameters and three different values of $Y_6$.
Observation of LNV is only possible, if $\Gamma(S_i^{\pm\pm}\to
l^{\pm}l^{\pm})$ is of similar order than $\Gamma(S_i^{\pm\pm}\to
W^{\pm}W^{\pm})$, since both final states are needed to establish that
LNV is indeed taking place. One can see from the figure that equality
of partial widths is possible for different choices of
parameters. However, since the decay to two charged leptons is
proportional to (the square of) a Yukawa coupling that is not fixed by
our neutrino mass fit, the relative ratio of branching ratios can not
be predicted from current data.

Similarly, also for all other decays to LNV final states, the two
competing final states have to have similar branching ratios. Fig.
(\ref{fig:LNV}) to the right show results for the decay of $\eta_3^{4+}$ of the
triplet model. Depending on the parameter $\mu_1$ equality of
branching ratio can occur in a large range of values of the parameter
$\lambda_2$. Note that the rate of LNV final states is not suppressed
by the smallness of neutrino masses. Neutrino masses require the
product of ${\cal F}\times Y_1Y_2$ to be small, see
eq. (\ref{eq:mnu2}).  For a fixed neutrino mass, smaller values of
$\mu_1\lambda_2$ require larger Yukawa couplings
$Y_1Y_2Y_3$. Depending on the ratio between $\mu_1\lambda_2$ and
$Y_1Y_2Y_3$, either the final state $4l$ or the final state $2l+2W$
can dominate. Whether LNV rates are observable, therefore, does not
depend so much on absolute values of some (supposedly small)
parameters, but on certain ratios of these parameters.

Table \ref{Tablelnv} is ordered with respect to increasing
multiplicity of the final state. Note that, as discussed in the last
subsection, cross sections at the LHC increase with electric charge
and decrease (strongly) with increasing mass. Which of the possible
signals has the largest rate, can not be predicted because of the
unknown mass spectrum. However, if the different members of the scalar
(or fermion) multiplets have similar masses, final states with larger
multiplicities have actually larger rates at the LHC. Since large
multiplicity final states also have lower backgrounds, searches for
such states should give stronger bounds.

The last column in table \ref{Tablelnv} gives our estimate for the
reach of the LHC. The numbers for the mass reach quoted in that column
are simply based on the cross section calculation, discussed in the
last subsection. Since in particular for the high multiplicity final
states we expect no SM backgrounds, we simply take the cross section for
which 3 events for a luminosity of 300 $fb^{-1}$ are produced as
the approximate limit, that maybe achieved in a dedicated search.
In fact, with supposedly no backgrounds even slightly lower masses than
those quoted in the table would lead to 5 or more events, sufficient
for a discovery.

On the other hand, our calculation does not include any cuts and
thus, should be taken only as a rough estimate. In particular,
for the simpler signal $ pp \rightarrow l^{+} l^{+} W^- W^- $, the
number given in the table should be taken with a grain of salt. 
Currently for dilepton searches with luminosity of 36 $fb^{-1}$ there
are no background events in the bins above $1$ TeV in the invariant
mass distribution $m (ll)$, see \cite{ATLAS:2017iqw}.  This in turns
implies for a luminosity of 300 $fb^{-1}$ in the most pessimistic case
an upper limit of roughly 8 background events for the signal $ pp
\rightarrow l^{+} l^{+} W^- W^-$. Our estimate of 3 signal events
would then correspond  only a 1 $\sigma$ c.l. limit. 

We mention that the final state with 2 $l$ and $6$ $W$'s and LNV
signals with 10 or more particles are also possible in $d=7$ 1-loop
models, but do not occur within our two example models. This is simply
due to the fact that scalars or fermions with 5 units of charge are
needed for such states. Thus, such signals can appear in versions of
the $d=7$ 1-loop type models, that include larger $SU(2)_L$
representations, such as quintuplets, or with particles with
a larger hypercharge.

Finally, the table considers only ``symmetric'' LNV final states.
Here, by symmetric we define that both branches of the decay contain
the same number of final states particles. For example, for the
quadruplet model, we have included the LNV signal with "symmetric"
final states $ p p \rightarrow \chi_2^{3+} \chi_2^{3-},
\chi_2^{3+} \rightarrow l^{+} l^{+} l^{+} , \chi_2^{3-}
\rightarrow W^- W^- l^{-} $, but we have not considered the possible
LNV signal with asymmetric final states $ p p \rightarrow \chi_2^{++}
\chi_2^{- - }, \chi_2^{++} \rightarrow l^{+} W^+ , \chi_2^{- - }
\rightarrow W^- W^- W^- l^{+} $. The reason for this choice is
simply that we consider ``asymmetric'' LNV signals, although in
principle possible,  are less likely to occur. This can be understood
simply from phase space considerations: A two-body final state
has a prefactor of $\frac{1}{8 \pi}$ in the partial width, while
a four-body phase space is smaller by a factor $3072 \pi^4$. 
Naturally one than expects that the ratio of branching ratios
for these asymmetric cases is never close to one, unless
there is a corresponding hierarchy in the couplings involved.

Decay widths for the lightest particle in our models are often very
small numerically. This opens up the possibility that some particle decays
might occur with a displaced vertex. Displaced vertices are more
likely to occur in the triplet model, so we concentrate in our
discussion on this case. The two-body decay width of $\eta_1^{++}$
is estimated in eq. (\ref{eq:GamPP}). For the decay of $\eta_3^{3+}$, 
assuming $\eta_3^{3+}$ is the lightest particle, one can
estimate:
\begin{equation}\label{eq:Gam3P}
  \Gamma(\eta_3^{+++} \to W^+l^+l^+) \sim \frac{1}{32\pi^2}
  \Big(\frac{\mu_1}{m_{\eta_2^{++}}^2}\Big)^2
  \frac{m_{\eta_3^{++}}^3}{m_{\eta_1^{++}}}\theta_{\eta_1\eta_2}^2
    \Gamma(\eta_1^{++}\to l^+l^+).
\end{equation}
Here, $\theta_{\eta_1\eta_2}$ is the mixing angle between the states
$\eta_1$ and $\eta_2$. Eq. (\ref{eq:Gam3P}) contains three parameters
related to the smallness of the observed neutrino masses: $\mu_1$,
$\theta_{\eta_1\eta_2}$ and $Y_1$.  Assuming all mass parameters
roughly equal $\mu_1\simeq m_{\eta_3^{++}} \simeq
m_{\eta_2^{++}}\simeq m_{\eta_1^{++}}=M$ this leads to the estimate:
\begin{equation}\label{eq:Len3P}
  L_0(\eta_3^{3+} \to W^+l^+l^+) \sim 0.3 
  \Big(\frac{10^{-1}}{\theta_{\eta_1\eta_2}}\Big)^2
  \Big(\frac{10^{-2}}{|Y_1|}\Big)^2
  \Big(\frac{10^{-2}}{|Y_4|}\Big)^2
  \Big(\frac{m_{\psi}}{\rm TeV}\Big)^2
  \Big(\frac{\rm TeV}{M}\Big) {\rm mm} .
\end{equation}
Here, the choice for the Yukawa couplings being order $10^{-2}$ is
motivated by the upper limits on the CLFV branching ratios, discussed
in the last section.  Eq. (\ref{eq:Len3P}) represents only a very rough
estimate, but it is worth pointing out that more stringent upper
limits from charged LFV would result in smaller values for the Yukawa
couplings, leading to correspondingly large decay lengths.  Note also
that smaller values of $\mu_1$ would lead to quadratically large
lengths. Eq. (\ref{eq:Len3P}) shows that displaced vertices can
occur easily in the decay of $\eta_3^{3+}$. 

Similarly, one can estimate roughly the order of magnitude of the decay
length for $\eta_3^{4+}$. The result is:
\begin{equation}\label{eq:Len4P}
  L_0(\eta_3^{4+} \to W^+W^+l^+l^+) \sim 4
  \Big(\frac{1}{\lambda_2}\Big)^2
  \Big(\frac{10^{-2}}{|Y_1|}\Big)^2
  \Big(\frac{10^{-2}}{|Y_4|}\Big)^2
  \Big(\frac{m_{\psi}}{\rm TeV}\Big)^2
  \Big(\frac{\rm TeV}{M}\Big) {\rm cm} .
\end{equation}
The width of $\eta_3^{4+}$ is smaller than the corresponding one for
$\eta_3^{3+}$ due to the phase space suppression for a 4-body final
state. Eq. (\ref{eq:Len4P}) shows that within the triplet model a
displaced vertex for the decay of $\eta_3^{4+}$ is actually expected.

\section{Discussion and conclusions\label{Sect:Con}}

In this paper we have discussed the phenomenology of $d=7$ 1-loop
neutrino mass models. Models in this class are far from the simplest
variants of BSM models that can fit existing neutrino data, but are
interesting in their own right, since they predict that new physics
must exist below roughly 2 TeV. If neutrino masses were indeed generated
by one of the models in this class, one can thus expect that the
LHC will find signatures of new resonances. Searches for doubly
charged scalars and multi-lepton final states already put some
bounds on these models. However, for the most interesting aspect
of $d=7$ 1-loop models, namely lepton number violating final
states, no LHC search exists so far. In particular, final states
with large multiplicites are predicted to occur (multiple $W$ and
multiple leptons) for which we expect standard model backgrounds
to be negligible.

In our discussion, we have limited ourselves to just two simple
example models. Our motivation to do so is that all $d=7$ 1-loop
neutrino mass models, which are genuine in the sense that they give
the leading contribution to neutrino mass without invoking new
symmetries, predict similar LHC signals. The two models which we
considered have either a $SU(2)_L$ triplet or a quadruplet as the
largest representations. Other $d=7$ models will contain even larger
$SU(2)_L$ multiplets and thus also particles with multiple electric
charges, to which very similar constraints than those analysed here
will apply. 

Finally, we mention that there exist variants of $d=7$ 1-loop models,
in which the internal scalars and fermions carry non-trivial colour
charges. These variants are not fully covered by our analysis. While
the neutrino mass fit and the constraints from LFV searches will be
qualitatively very similar to what we have discussed here, additional
color factors in the calculations will lead to some quantitative
changes. The resulting bounds will, in general be somewhat more
stringent than the numbers we give in this paper. More important,
however, are the changes in the LHC phenomenology. For example, in the
colour-singlet models, which we analyzed in this paper, the lightest
doubly charged scalar will decay to two charged leptons. In the
coloured variants of the model, the corresponding lightest scalar will
behave like a leptoquark, decaying to $l+j$, instead. Thus, different
LHC searches will apply to the coloured $d=7$ models.  More
interesting, however, is that for coloured models also the LNV
final states, which we discussed, will change, since at the end
of the decay chain instead of two charged lepton, one lepton plus
jet will appear. Although this variety of signals will be interesting
in their own rights, we have concentraged here on the colour
singlet variants of the model, because di-leptons are cleaner
(and thus more easy to probe) in the challenging experimental
environment that is the LHC.

\medskip
\centerline{\bf Acknowledgements}

\medskip
This work was supported by the Spanish MICINN grants FPA2014-58183-P, FPU15/03158 (MECD) and PROMETEOII/2014/084 (Generalitat Valenciana). J.C.H. is supported by Chile grants Fondecyt No. 1161463,
Conicyt  PIA/ACT 1406 and Basal FB0821. 


\bibliography{references_0nubb}
\bibliographystyle{h-physrev5}

\end{document}